# Atomic-scale observation and manipulation of plaquette antiferromagnetic order in iron-based superconductor

Seokhwan Choi[1], Won-Jun Jang[1,2], Jong Mok Ok[3], Hyun Woo Choi[1], Hyun-Jung Lee[3], Se-Jong Kahng[4], Young Kuk[5], Ja-Yong Koo[6], SungBin Lee[1], Sang-Wook Cheong[7], Yunkyu Bang[8], Jun Sung Kim[3], Jhinhwan Lee[1,2*]

[1] Department of Physics, Korea Advanced Institute of Science and Technology, Daejeon 34141, Korea
[2] Center for Axion and Precision Physics, Institute of Basic Science, Daejeon 34141, Korea
[3] Department of Physics, Pohang University of Science and Technology, Pohang 37673, Korea
[4] Department of Physics, Korea University, Seoul 136-713, Korea
[5] Department of Physics and Astronomy, Seoul National University, Seoul 151-747, Korea
[6] Korea Research Institute of Standards and Science, P.O. Box 102, Yuseong, Daejeon 305-600, Korea
[7] Rutgers Center for Emergent Materials and Department of Physics and Astronomy, Rutgers University, Piscataway, New Jersey 08854, USA
[8] Department of Physics, Chonnam National University, Kwangju 500-757, Korea

[*]e-mail: jhinhwan@kaist.ac.kr

(Originally prepared: June 8, 2016)
(Updated: August 15, 2016)

**Abstract**

The symmetry requirement and the origin of magnetic orders coexisting with superconductivity have been strongly debated issues of iron-based superconductors (FeSCs). Observation of $C_4$-symmetric antiferromagnetism in violation of the inter-band nesting condition of spin-density waves in superconducting ground state will require significant change in our understanding of the mechanism of FeSC. The superconducting material $Sr_2VO_3FeAs$, a bulk version of monolayer FeSC in contact with a perovskite layer with its magnetism ($T_N \sim 50$ K) and superconductivity ($T_c \sim 37$ K) coexisting at parent state, has no reported structural orthorhombic distortion and thus makes a perfect system to look for theoretically expected $C_4$ magnetisms[1-3]. Based on variable temperature spin-polarized scanning tunneling microscopy (SPSTM) with newly discovered imaging mechanism that removes the static surface reconstruction (SR) pattern by fluctuating it rapidly with spin-polarized tunneling current, we could visualize underlying $C_4$ symmetric ($2 \times 2$) magnetic domains and its phase domain walls. We find that this magnetic order is perfectly consistent with the plaquette antiferromagnetic order in tetragonal Fe spin lattice expected from theories based on the Heisenberg exchange interaction of local Fe moments and the quantum order by disorder[4]. The inconsistency of its modulation $Q$ vectors from the nesting condition also implies that the nesting-based $C_2$ symmetric magnetism is not a unique prerequisite of high-$T_c$ FeSC. Furthermore, the plaquette antiferromagnetic domain wall dynamics under the influence of small spin torque effect of spin-polarized tunneling current are shown to be consistent with theoretical simulation based on the extended Landau-Lifshitz-Gilbert equation.

[1]   Iron-based superconductors (FeSCs) have shown wide range of intriguing phenomena related to the coexistence of magnetism and superconductivity below the superconducting transition temperature ($T_c$)[5-7]. Although understanding of their detailed interplay mechanism is still in debate, certain magnetic orders or magnetic fluctuations seem very crucial in realizing coexistent superconductivity[6]. This demands us to figure out the exact magnetic ground states and their fluctuations in FeSCs starting from the simplest form of FeSC layers with little magnetoelastic couplings to structural deformation or inter-FeSC-layer couplings. In understanding the nature of magnetism in FeSCs, two scenarios (or their combinations) have been widely studied; the nesting picture of spin density waves and the interacting localized moment picture[8,9]. In the former nesting-based scenario, itinerant Fe electrons satisfying nesting condition between the Γ and the M (or X) bands lead to spin density wave-based magnetic ground states often with $C_2$ symmetric stripe orders and sometimes accompanies charge density wave or nonuniform superconducting states[10-15]. In the latter scenario, however, Fe local moments can interact via short ranged exchange interactions or long ranged Ruderman-Kittel-Kasuya-Yoshida (RKKY) interactions in the existence of itinerant electrons resulting in magnetic orders explained by effective Heisenberg Hamiltonians.

[2]   In the localized moment scenario, the exchange interactions between the Fe spins $\vec{S_i}$ can be typically described by $J_1$-$J_2$-$J_3$ Heisenberg model (or $J_1$-$J_2$-$J_3$-$K$ with $K$ as a coefficient describing the tendency towards the nearest-neighbor collinearity or the existence of easy axis/plane)[4,6,7,19-21]. Especially in the regime of $J_3 > \frac{|J_2|}{2}$ and $J_3^2 \gtrsim SJ_1^2$, where $S$ is the spin magnitude, it is understood that the unique 'plaquette antiferromagnetic order' with $C_4$ symmetry is selected by quantum fluctuations in a mechanism often referred as quantum order by disorder[19,21]. This is discussed in detail in the Supplementary Section 1. This plaquette order (a.k.a. orthogonal double stripe (ODS) order) is described by two orthogonal spin modulation

wave vectors $\boldsymbol{Q} = \left(\frac{\pi}{2}, \frac{\pi}{2}\right)_{\text{Fe}}$ and $\boldsymbol{Q}^* = \left(\frac{\pi}{2}, -\frac{\pi}{2}\right)_{\text{Fe}}$ in each of two Fe sublattices (The subscript 'Fe' implies a single Fe unit-cell basis whereas no subscript implies the full two-Fe unit-cell basis), and their wave vectors may not be consistent with the Fermi surface nesting wave vectors[4]. This plaquette antiferromagnetic order with $C_4$ symmetry, however, has never been observed directly, because strong lattice orthorhombicity of $C_2$ symmetry found in most FeSCs favors orders with the same symmetry due to magnetoelastic effects. For example, in the same parameter space suitable for plaquette order, a diagonal double stripe (DDS) order with $C_2$ symmetry may be preferred in the presence of strong $C_2$ lattice symmetry, where the two Fe sublattices have the same spin modulation wavevector $\boldsymbol{Q} = \left(\frac{\pi}{2}, \frac{\pi}{2}\right)_{\text{Fe}}$ (Ref. Supplementary Section 1). Such DDS orders can be classically stabilized with energy gain of $-O(|J_{2a} - J_{2b}|)$ due to the anisotropy $J_{2a} \neq J_{2b}$ associated with the lattice orthorhombicity and indeed has been observed in non-superconducting parent compound of $Fe_{1+y}Te$ with strong orthorhombic lattice distortion[23]. So far, a direct real-space evidence of local-moment-based magnetic order with any (either $C_4$ or $C_2$) symmetry coexisting with superconductivity in FeSCs has never been observed.

[3]    In the parent state superconductor $Sr_2VO_3FeAs$, it has been reported that the magnetism ($T_N \sim 50$ K) coexists with superconductivity ($T_c \sim 37$ K) in the absence of any orthorhombicity at all[1-3] and the inter-FeSC-layer spacing is larger than most other FeSCs. Therefore, it is an ideal system where the pure magnetic ground state and its quantum fluctuation effects can be studied without strong anisotropic magnetoelastic strains or strong inter-FeSC-layer coupling. Furthermore, its crystal structure shows that every FeAs monolayer is sandwiched by two perovskite $Sr_2VO_3$ monolayers as shown in Fig. 1a, making it the only currently known bulk version of the monolayer FeSC in contact with perovskite layer with fascinating property of interfacial phonon-assisted $T_c$ enhancement[24-26].

[4]   For real-space magnetic imaging on FeSCs, we have developed a novel technique of spin-polarized scanning tunneling microscopy (SPSTM) with antiferromagnetic Cr-cluster tip created *in situ* on a Cr surface and performed a comparative study with W tip and Cr-cluster tip. The 4.6 K STM topographic images of cold-cleaved (~10 K) surfaces taken with non-spin-polarized W tip in Fig. 1d and Figs. S4 show small randomly oriented (2×n) (n=3,5,8) domains of quasi-$C_2$-symmetric atomic corrugations, consistent with surface reconstruction (SR) with no clear signature of bulk orthorhombicity (to be discussed below in terms of Fig. 2 and 3). On the other hand, our SPSTM images with in-plane-polarized Cr-cluster tip, beyond a small bias threshold (~40 meV, ~25 pA), show clear $C_4$ symmetric (2×2) magnetic unit cells with intra-unit-cell topographic modulations consistent with Fe-lattice plaquette antiferromagnetic order as shown in Fig. 1e-f without any signature of SR seen in Fig. 1d. This is a surprising phenomenon later found to be due to the fast SR fluctuations driven by the spin-polarized tunneling current, which effectively flattens the SR by time averaging so that only the magnetic patterns by SPSTM mechanism are clearly seen in Fig. 1e-1f. Another important point for this material is that we have well-defined vertical tunneling path consisting of O-V-As ions to the FeAs layer, because in the top SrO layer only oxygen atoms has significant density of states[27] and, as a result, any magnetic signal of Fe-layer observed at the top layer oxygen should be the average of the four neighboring Fe spins connected to the As ions in each vertical O-V-As tunneling path. Therefore the observed (2×2) pattern with three groups of apparent height levels is perfectly consistent with the plaquette order in the Fe lattice. The spin modulation wave vectors $\boldsymbol{Q} = \left(\frac{\pi}{2}, \frac{\pi}{2}\right)_{Fe}$   and $\boldsymbol{Q}^* = \left(\frac{\pi}{2}, -\frac{\pi}{2}\right)_{Fe}$   are visible in the Fourier transformed q-space image (the inset of Fig. 1e) and are shown as white arrows in Fig. 1c where the ARPES-based Fermi surface[28] is shown as dark curves. The inconsistency between the magnetic order

wave vectors and the inter-band nesting vectors is clearly visible which is a clear signature of interacting local moment picture.

[5]   To understand the physics of SPSTM imaging on a FeSC layer through a perovskite layer, we performed a comparative study of bias-dependent measurements using W tip and Cr tip. In case of W tip, as shown in Fig. 2a-e and Fig. S8, we found that the SR patterns start to change at bias beyond $V_W \sim 280$ meV and the fluctuation becomes so rapid above 400 meV that the surface starts to appear essentially flat as a result of time-averaging of the fluctuations. As we go back to the low bias, the SR fluctuation is frozen leaving a state of randomized SR. On the other hand, in case of Cr tip, as shown in Fig. 2f-j and Fig. S9, the surface looks indistinguishable to the W tip case at low bias near 10 meV but the surface patterns start to change at a much lower threshold bias near 40 meV, revealing the (2×2) domain structure and its phase domain walls without the signature of the original SR. (At this bias condition with low junction conductance, the (1x1) lattice modulations appear weak compared with the high junction conductance case of Fig. 1e-f.) As we go back to the low bias voltage near 10 meV, the surface fluctuation rate is decreased but not completely frozen leading to the individual SR fluctuation visible as horizontal streaks. This implies that, in case of Cr tip, the surface changes its characteristics beyond the initial low bias threshold $V_{Cr1} \sim 40$ meV such that the activation threshold for the SR fluctuation is significantly lowered $V_{Cr2} < 10$ meV. Therefore, the Cr tip imaging on $Sr_2VO_3FeAs$ with bias slightly higher than the initial activation threshold will start to experience rapid SR fluctuations, resulting in no sign of static SR patterns since the time average of fast fluctuating SR should be effectively flat across the surface. Therefore, the magnetic contrast of SPSTM stands out in the flat background as shown in Fig. 1e-f, and the possibility of the (2×2) patterns being another form of static SR is ruled out. From the fact that the spin-polarized Cr tip has much lower threshold for SR fluctuation compared with the W tip,

we can deduce that the magnetic excitation channel through the V magnetic moment coupled to the SR is playing an important role in the SR fluctuations under the Cr tip.

[6]   To visualize the temperature-dependence of the symmetries of the lattice and the magnetic orders in this interesting FeSC, we developed variable temperature SPSTM technique with temperature range spanning the superconducting critical temperature ($T_c \sim 37$ K), the magnetic transition temperature ($T_N \sim 50$ K) and a higher transition temperature for a poorly understood order parameter ($T^* \sim 150$ K)[3]. Even though there has been no single consensus on the nature of magnetism in this material[3,29,30], there have been reports that the magnetic transition at $T_N$ is due to the Fe moment ordering instead of V[3,30]. The W-tip topographs taken at various temperatures during the cooling process after warming up the sample to 180 K, show the field of view filled with a single domain of quasi-$C_2$-symmetric SR below 150 K down to 4.6 K, as shown in Fig. 3a-e which may be understood either as the annealing-induced domain growth of the $C_2$ symmetric SR even in the absence of the bulk orthorhombicity or as a signature of an unknown $C_2$ symmetric order parameter below 150 K. On the other hand, the Cr-tip-based SPSTM topographs shows $C_4$ symmtery above 150 K and changes into $C_2$ symmetry below 150 K down to 60 K. From 60 K down to 40 K, the domains of $C_4$ symmetric plaquette antiferromagnetic order (visible as the bright (2×2) patches) slowly grows to cover the entire field of view, as shown in Fig. 3f-j. From 40 K down to 4.6 K, the long range $C_4$ plaquette order with well defined phase domain walls (pDWs) persists.

[7]   This plaquette order is distinct from the $C_4$ magnetic orders in other FeSCs induced from inter-band nesting reported by neutron scattering measurement[31,32]. The current school of thought states that nesting-based magnetism harbors FeSC and in most cases $C_2$ magnetism is favored over $C_4$ magnetism[33,34]. However, our temperature dependent SPSTM results show that Fe local moment-based $C_4$ plaquette magnetism may stably harbor FeSC with $T_c \sim 37$ K

and give a single clear proof that the $C_2$ symmetric nesting-based magnetism is not a unique prerequisite of high-$T_c$ FeSC.

[8]    So far we have demonstrated that the Fe magnetic moments forming the plaquette order are highly stable as shown in Fig. 1e-f, even under the fast SR fluctuations caused by the strong spin-polarized current. However, the Fe spins near the phase domain walls (pDWs) are intrinsically metastable due to competition of the two neighboring phase domains. Therefore, we can expect that the spin-torque effect of the spin-polarized current can result in pDW motions. The Figs 4a,b show a large area SPSTM topograph (a) and its pDWs detected by spatial lock-in technique (b) (Ref. Supplementary Section 6). Figure 4c is taken immediately after Fig. 4a,b which reveals that there are plenty of pDW changes due to the spin-torque effect by the spin-polarized tunneling current scanned over the whole field-of-view (Ref. Supplementary Section 7). We can classify the pDWs by three colors (purple, red, blue) in terms of the relative phase shift of the domains separated by them and also by three Greek letters ($\alpha, \beta, \gamma$) in terms of the symmetry of pDW spin configurations, as shown in Figs 4d-i. The minimum energy spin configurations in all three kinds of pDWs show spin spiral structures with its pDW width up to several Fe lattice constants as shown in the Fig. 4j-l. The α-pDW has alternating two distinct 1D spiral magnetic DWs lying perpendicular to the pDW, i.e. antiferromagnetic spiral DWs (long green rectangles) and ferromagnetic spiral DWs (long yellow rectangles). Other pDWs have only one kind of spiral DWs, i.e. 1D antiferromagnetic spiral DWs for β-pDWs and 1D ferromagnetic spiral DWs for γ-pDWs.

[9]    In order to understand the mechanism of pDW motion under the influence of the spin torque effect of the spin-polarized tip current, we performed the following theoretical experiments. We assumed that the spin torque effect by the tunneling current is ferromagnetic and is applied over a small range (a few Fe lattice constants) around the tip. The spins

interacting with themselves (through the $J_1$-$J_2$-$J_3$-$K$ Heisenberg Hamiltonian) and with the tunneling current (through the ferromagnetic spin-torque effect) can be modelled by the extended Landau-Lifshitz-Gilbert equation in the form of $\frac{\partial \vec{s}_i}{\partial t} = -\frac{\gamma}{(1+\alpha^2)\mu_S} \vec{S}_i \times \left[\vec{H}_i + \alpha\left(\vec{S}_i \times (\vec{H}_i + \vec{T}_i)\right)\right]$[35] (Supplementary Section 8).

[10]   The effect of tip scanning on the pDWs by spin torque effect of the spin polarized tunneling current is visualized in the sequential images shown in Fig. 5a-h. The common behaviors of all different pDWs in response to the spin-polarized tunneling current is that the spin torque effect tends to pull the pDWs toward the tip whenever the pDW is within the effective interaction range of the tip (shown as bright area around the tip position). This can be understood as the zone of ferromagnetism generated near the tip due to the spin-torque effect confuses and weakens the antiferromagnetic interactions among the Fe moments in the domain near the tip, and lets the domain in the opposite side of the pDW to advance toward the tip breaking the balance between the two domains. A side effect is that a small segment of the pDW tends to be left in the last state of being pulled toward the direction the tip was moving away and its effect is strong especially for β- and γ-pDWs with smaller number of spin changes required for pDW shift. The larger number of spin changes required for the motion of α-pDW makes the α-pDWs kinetically less susceptible to the spin-torque effect of scanning spin-polarized tip. This is in good agreement with the experimental observation that the β- and γ-pDWs are more mobile under the scanning tip in comparison with α-pDWs. Figure 5i-p shows a more realistic simulation near a phase domain island near a pDW which is modelled after the pDW configuration in the green square in Fig.4b and reveals how the abrupt change of the phase of the small domain island as well as the dragging of the surrounding pDW contribute to the complete disappearance of the localized phase domain, as seen in the green square in Fig.4c.

[11]    In conclusion, we observed $C_4$-symmetric plaquette antiferromagnetic order in a parent state iron-based superconductor $Sr_2VO_3FeAs$, currently the unique self-assembled bulk version of a monolayer FeSC on a perovskite substrate, using SPSTM measurement with antiferromagnetic Cr-cluster tip. The rapid fluctuations of the SR induced by the spin-polarized current of the Cr tip effectively erase the static SR out of the magnetic contrast image. The significantly lower bias threshold for SR fluctuation in case of Cr tip compared with W tip case implies that the magnetic channel of excitation linked to the V spin degrees of freedom is more efficient than the charge channel in terms of the SR fluctuations. Our variable temperature SPSTM measurement shows the appearance of $C_4$ magnetic symmetry from $T_M \sim 50$ K (just above $T_c \sim 37$ K) and all the way down to 4.6 K, beyond a small threshold bias. Also, the wave vectors for the (2×2) magnetic order we found are inconsistent with the inter-band nesting condition. The observation of stable plaquette order even under the rapid SR fluctuations and the spin-torque effect from the spin-polarized tunneling current implies that the plaquette order can be a universally occurring ground state due to quantum fluctuations and order-by-disorder principle. Our spin-dynamics simulations of pDW motions under the spin-polarized tunneling current is consistent with our experimental observations and explain how the spin-toque effect due to the spin-polarized tunneling current can induce the motions of pDWs, as well as the abrupt phase reversal of small domain islands, by breaking the delicate balance between competing antiferromagnetic domains with local injection of ferromagnetic interactions. This may open a possible route to controlling complex $C_4$ antiferromagnetic domains with spin-polarized current injection.

**References**

1. Zhu, X. *et al*. Transition of stoichiometric $Sr_2VO_3FeAs$ to a superconducting state at 37.2 K. *Phys. Rev. B* **79**, 220512 (2009).

2. Hummel, F., Su, Y., Senyshyn, A. & Johrendt, D. Weak magnetism and the Mott state of vanadium in superconducting $Sr_2VO_3FeAs$. *Phys. Rev. B* **88,** 144517 (2013).

3. Ok., J. M., Baek, S. H. & Kim, J. S., in preparation.

4. Ducatman, S., Perkins, N. B. & Chubukov, A. Magnetism in parent iron chalcogenides: Quantum fluctuations select plaquette order. *Phys. Rev. Lett.* **109**, 157206 (2012).

5. Hosono, H. & Kuroki, K. Iron-based superconductors: Current status of materials and pairing mechanism. *Physica C* **514**, 399–422 (2015).

6. Dai, P. Antiferromagnetic order and spin dynamics in iron-based superconductors. *Rev. Mod. Phys*. **87**, 3 (2015).

7. Johnston, D. C. The puzzle of high temperature superconductivity in layered iron pnictides and chalcogenides. *Adv. Phys.* **59**, 803-1061 (2010).

8. Paglione, J. & Greene, R. L. High-temperature superconductivity in iron-based materials. *Nat. Phys.* **6**, 645–658 (2010).

9. Dai, P., Hu, J. & Dagotto, E. Magnetism and its microscopic origin in iron-based high-temperature superconductors. *Nat. Phys.* **8**, 709-718 (2012).

10. Ma, F. and Lu, Z.-Y. Iron-based layered compound LaFeAsO is an antiferromagnetic semimetal. *Phys. Rev. B* **78**, 033111 (2008).

11. Dong, J. *et al*. Competing orders and spin-density-wave instability in $La(O_{1-x}F_x)FeAs$. *Europhys. Lett.* **83**, 27006 (2008).

12. Vorontsov, A. B., Vavilov, M. G. and Chubukov, A. V. Superconductivity and spin-density waves in multiband metals. *Phys. Rev. B* **81**, 174538 (2010).

13. Eremin, I. and Chubukov, A. V. Magnetic degeneracy and hidden metallicity of the spin-density-wave state in ferropnictides. *Phys. Rev. B* **81**, 024511 (2010).

14. Cvetkovic, V. and Tesanovic, Z. Valley density-wave and multiband superconductivity in iron-based pnictide superconductors. *Phys. Rev. B* **80**, 024512 (2009).

15. Balatsky, A. V., Basov, D. N. and Zhu, J. -X. Induction of charge density waves by spin density waves in iron-based superconductors. *Phys. Rev. B* **82**, 144522 (2010).

16. Ma, F., Ji, W., Hu, J., Lu, Z.-Y. & Xiang, T. First-Principles Calculations of the Electronic Structure of Tetragonal $\alpha$-FeTe and $\alpha$-FeSe Crystals: Evidence for a Bicollinear Antiferromagnetic Order. *Phys. Rev. Lett.* **102**, 177003 (2009).

17. Xia, Y. *et al.* Fermi Surface Topology and Low-Lying Quasiparticle Dynamics of Parent $Fe_{1+x}Te/Se$ Superconductor. *Phys. Rev. Lett.* **103**, 037002 (2009).

18. Yi, M. *et al.* Observation of universal strong orbital-dependent correlation effects in iron chalcogenides. *Nat. Comm.* **6**, 7777 (2015).

19. Chandra, P., Coleman, P. & Larkin, A.I. Ising Transition in Frustrated Heisenberg Models. *Phys. Rev. Lett.* **64**, 88-91 (1988).

20. Chubukov, A. First-order transition in frustrated quantum antiferromagnets. *Phys. Rev. B* **44**, 392 (1991).

21. Glasbrenner, J. K., Mazin, I. I., Jeschke, H. O., Hirschfeld, P. J., Fernandes, R. M., Valentí, R., Effect of magnetic frustration on nematicity and superconductivity in iron chalcogenides. *Nat. Phys.* **11**, 953–958 (2015).

22. Villain, J., Bidaux, R., Carton, J.-P., Conte, R. Order as an effect of disorder. *J. Physique* 41, 1263-1272 (1980).

23. Enayat, M. *et al*. Real-space imaging of the atomic-scale magnetic structure of $Fe_{1+y}Te$. *Science* **345**, 653-656 (2014).

24. Ge, J.-F. Superconductivity above 100 K in single-layer FeSe films on doped $SrTiO_3$. *Nat. Mater.* **14**, 285-289 (2015).

25. Fan, Q. *et al*. Plain *s*-wave superconductivity in single-layer FeSe on $SrTiO_3$ probed by scanning tunnelling microscopy. *Nat. Phys.* **11**, 946-952 (2015).

26. Lee, D.-H., What makes the $T_c$ of $FeSe/SrTiO_3$ so high? *Chin. Phys. B* **24**, 117405 (2015).

27. Lee, K.-W. & Pickett, W. E. $Sr_2VO_3FeAs$: A nanolayered bimetallic iron pnictide superconductor. *Europhys. Lett.* **89,** 57008 (2010).

28. Kim, Y.K., *et al*. Possible role of bonding angle and orbital mixing in iron pnictide superconductivity: Comparative electronic structure studies of LiFeAs and $Sr_2VO_3FeAs$. *Phys. Rev. B* **92**, 041116(R) (2015).

29. Tegel, M. *et al*. Possible magnetic order and suppression of superconductivity by V doping in $Sr_2VO_3FeAs$. *Phys. Rev. B* **82**, 140507 (2010).

30. Ueshima, K., *et al*. Magnetism and superconductivity in $Sr_2VFeAsO_3$ revealed by $^{75}As$- and $^{51}V$-NMR under elevated pressures. *Phys. Rev. B* **89**, 184506 (2014)

31. Avci, S. *et al*. Magnetically driven suppression of nematic order in an iron-based superconductor. *Nat. Comm.* **5**, 3854(2014).

32. Allred, J. M. *et al*. Double-Q spin-density wave in iron arsenide superconductors, *Nat. Phys.* **12**, 493–498 (2016).

33. Böhmer, A. E. *et al*. Superconductivity-induced re-entrance of the orthorhombic distortion in $Ba_{1-x}K_xFe_2As_2$, *Nat. Comm.* **6**, 7911 (2015).

34. Wang, L. *et al*. Complex phase diagram of $Ba_{1-x}Na_xFe_2As_2$: A multitude of phases striving for the electronic entropy. *Phys. Rev. B* **93**, 014514 (2016).

35. Wieser, R., Vedmedenko, E. Y. & Wiesendanger, R. Indirect control of antiferromagnetic domain walls with spin current. *Phys. Rev. Lett.* **106**, 067204 (2011).

**Acknowledgement**

The authors are thankful for the helpful discussions with A. Chubukov, C. Kim, J.J. Yu, M. Allan, T.M. Chuang, E.-A. Kim, A. Heinrich, W. Wu, J.H. Shim, A.T. Lee and Y.K. Semertzidis.

J.L. was supported by the Metrology Research Center Program funded by Korea Research Institute of Standards and Science (No. 2015-15011069), by the Pioneer Research Center Program through the National Research Foundation (NRF) (No. 2013M3C1A3064455), by the Basic Science Research Programs through the NRF (No. 2013R1A1A2010897 and No. 2012R1A1A2045919), and by Samsung Advanced Institute of Technology (SAIT).

S.W.C. was supported by the Gordon and Betty Moore Foundation's EPiQS Initiative through Grant GBMF4413 to the Rutgers Center for Emergent Materials.

J.S.K. was supported by the NRF through SRC Center for Topological Matter (No. 2011-0030046), and the Max Planck POSTECH/KOREA Research Initiative Program (No. 2011-0031558).

**Author Contributions**

Sample synthesis: J.M.O. and J.S.K.

Development of the SPSTM method: J.L.

STM experiment: S.C., J.L., H.W.C., and W.J.J.

Data analysis: J.L., S.C., S.W.C., H.W.C., W.J.J., S.B.L., and H.J.L.

All authors participated in writing the manuscript.

**Additional Information**

Correspondence and requests for materials should be addressed to J.L.

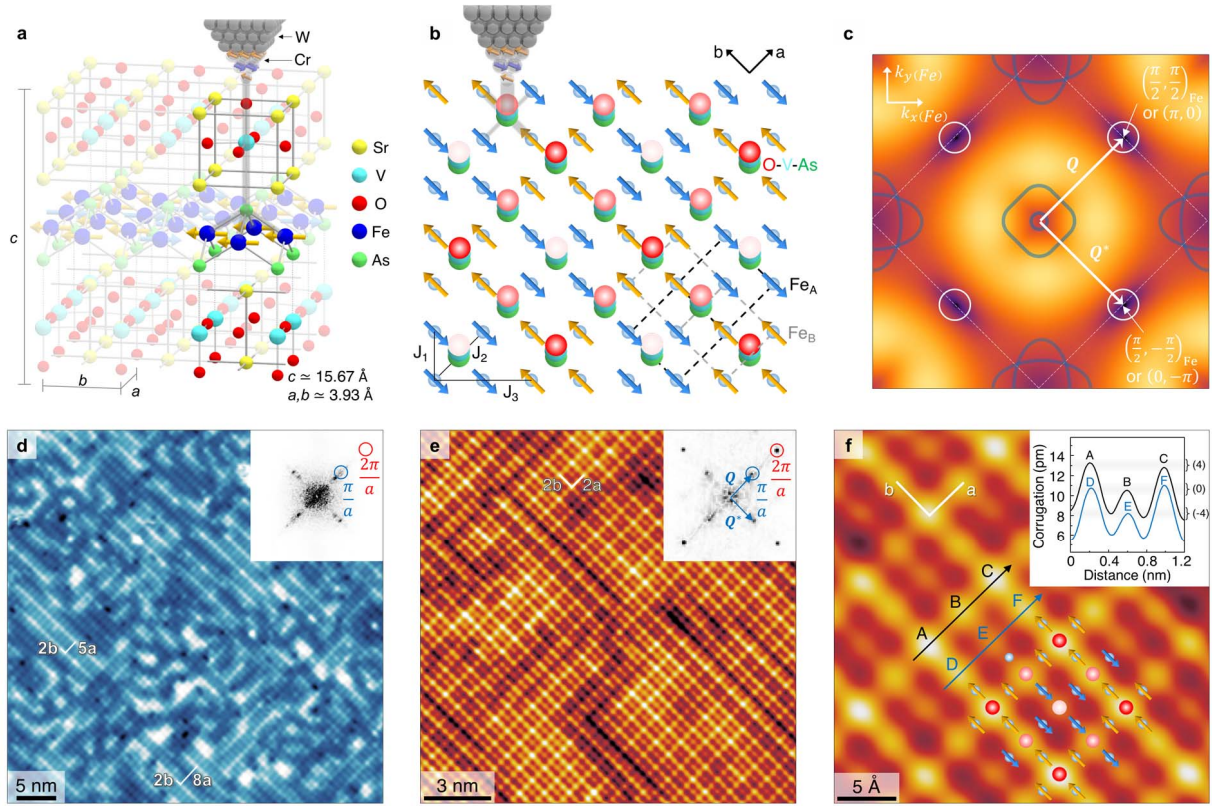

**Figure 1. The plaquette (ODS) order observed in Sr₂VO₃FeAs with spin-polarized STM. a**, A structural model of Sr₂VO₃FeAs lattice with the tunneling path through a vertical O-V-As atomic chain shown in gray. **b**, The spin structure of the Fe moments in the plaquette order and the apparent (2×2) magnetic unit cell in SPSTM. Each red dot represents the Oxygen at the top of each vertical O-V-As atomic chain, with the dot's intensity indicating the net spin-polarized signal, determined by the average of the four neighboring Fe spins relative to the Cr tip spin. **c**, The magnon dispersion and the two Q vectors from localized moment picture, for the long range plaquette spin order shown together with the ARPES-based Fermi surfaces (dark curves). The subscript 'Fe' implies a single Fe unit-cell basis whereas no subscript implies the full two-Fe unit-cell basis. **d**, A typical 4.6 K STM topograph of cold-cleaved (~10 K) surface taken with W tip at bias condition (60 meV, 80 pA) showing glassy $C_2$ SRs with random orientations. **e**, A spin-polarized STM topograph taken with in-plane polarized Cr tip at bias condition of (150 meV, 500 pA) showing $C_4$-symmetric (2×2) magnetic orders. The insets in **d** and **e** are respective $q$-space images. **f**, The magnified view of the (2×2) magnetic unit cells with a plaquette spin model overlayed. Inset shows topographic cross-sections along the black and blue arrows in **f**. The numbers marked on each peak group indicates the net spin component of four neighboring Fe spins parallel to the Cr tip spin.

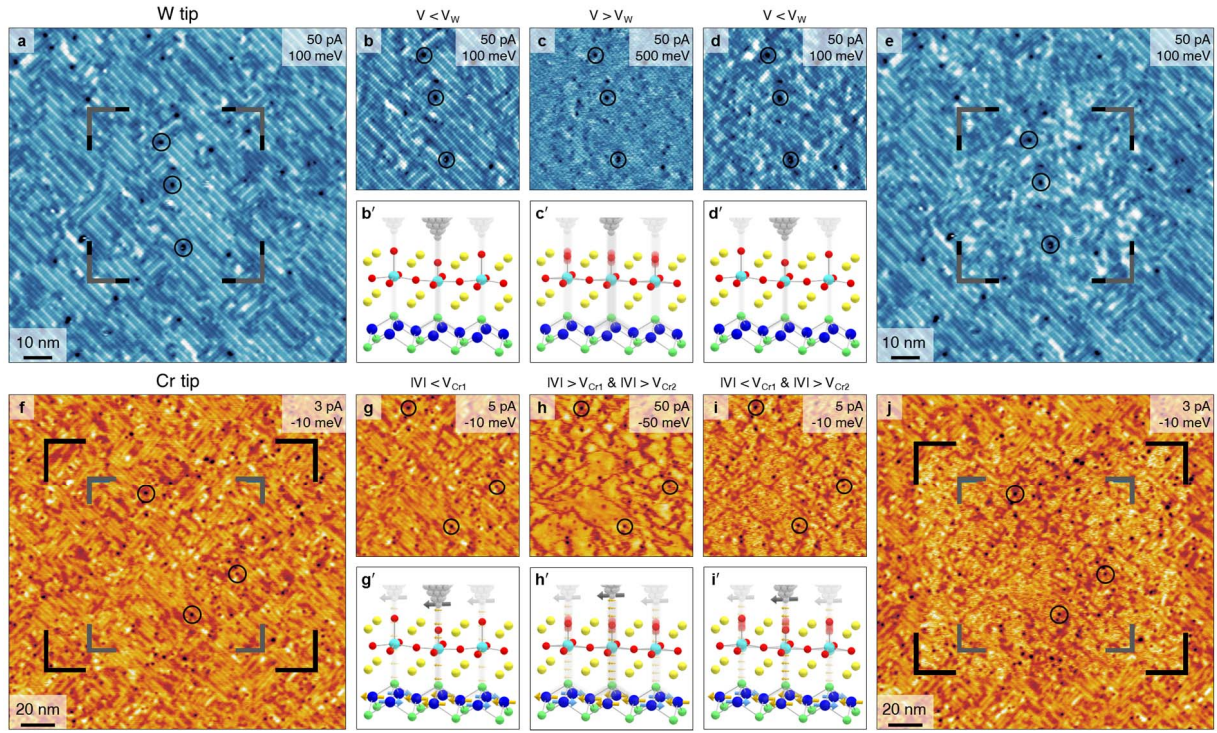

**Figure 2. The dichotomy between the W tip and the Cr tip in terms of the bias-dependent fluctuations of the SR. a-e,** In case of the non-spin-polarized W tip, there is a single high threshold ($V_W \sim 280$ mV) of the bias voltage above which the SR fluctuates rapidly resulting in an apparently flat surface (Ref. Supplementary Information Section 5). As we reduce the bias voltage back to the original value below $V_W$ (**d,e**), we find a strongly disordered SR patterns with modulation wavelengths similar to those found in **a**, mostly confined inside the region of high bias W tip scanning. In the large scale images **a, e, f** and **j**, the areas actually scanned with high bias conditions are bounded by the black corner markers while the areas displayed in **b-d** and **g-i** are bounded by the gray corner markers. **f-j,** In case of the spin-polarized Cr tip, only at very low bias condition (**f**) we can observe SR equivalent to those found in **a**. However, the initial bias voltage threshold for the SR fluctuation is much lower ($V_{Cr1} \sim 40$ mV) and just above it we immediately start to observe $C_4$ symmetric (2×2) domains and its phase domain walls (**h**), without any hint of the original SR in **f**. This apparent absence of surface fluctuation for $|V| > V_{Cr1}$ is because the spin-polarized-current-enhanced SR fluctuation is fast enough and time-averaged to a constant over the field of view (**h′**). This can be verified by reducing the bias condition (**i** and **j**) back to the very low bias used for **f** or **g**, where we start to see the individual SR fluctuation as the horizontal streaks. This implies that as soon as the initial threshold ($V_{Cr1}$) is reached and the surface is changed into fluctuating condition, we have a new threshold for SR fluctuation significantly reduced ($V_{Cr2} < 10$ mV) from the initial $V_{Cr1}$. The significantly lower thresholds of SR fluctuation in case of the spin-polarized tip shows the strong coupling of the SR in the top SrO layer and the spin degree of freedom of the V in the intermediate $VO_2$ layer. The atomic scale models for the STM/SPSTM topograph measurement conditions in **b-d** and **g-i** are shown in **b′-d′** and **g′-i′**.

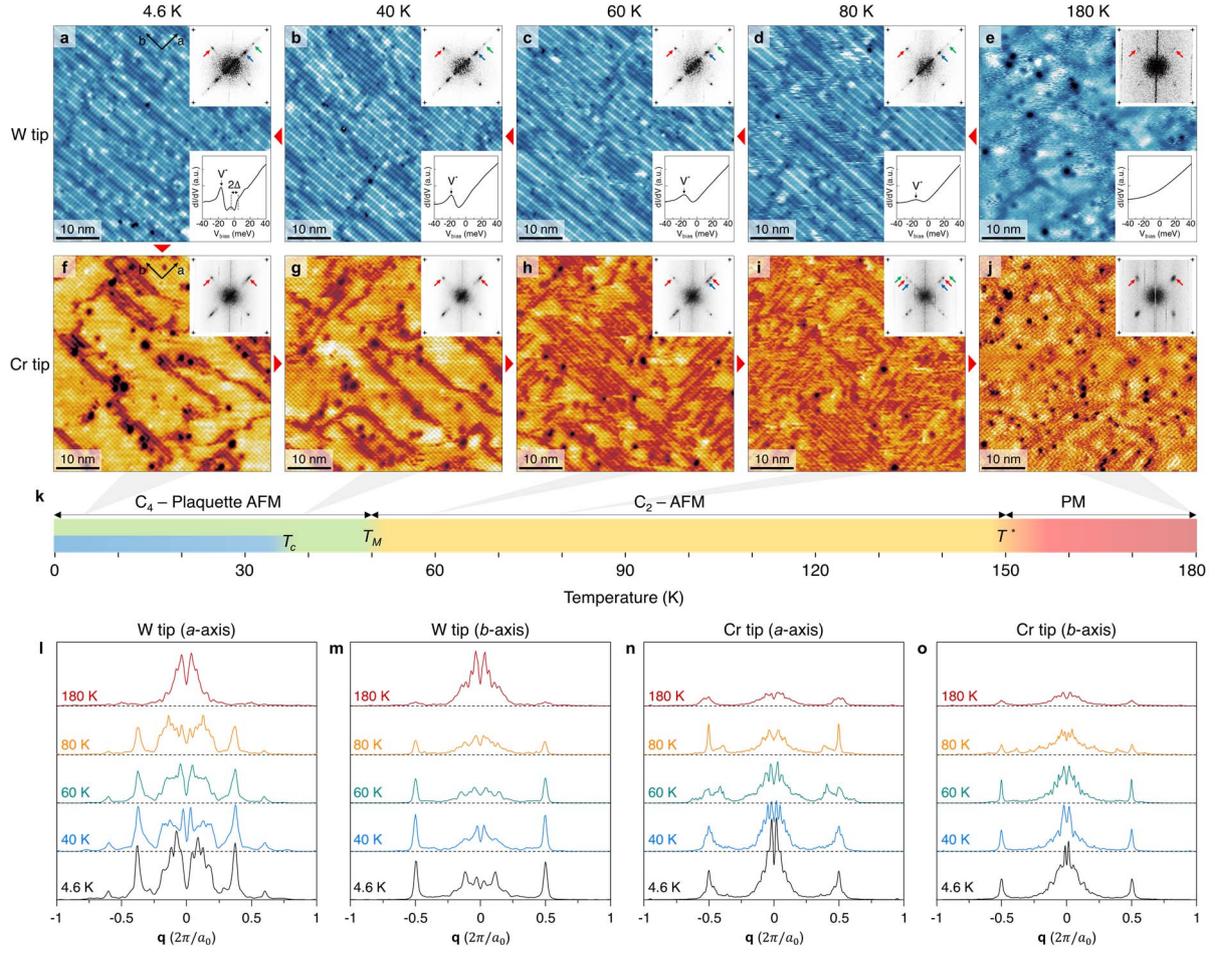

**Figure 3. Temperature-dependence of the magnetic domain structure in $Sr_2VO_3FeAs$.** The top row (**a-e**) shows normal topographs taken with W tip at bias condition of (50 meV, 100 pA) and the bottom row (**f-j**) shows SPSTM topographs taken with spin-polarized Cr tip at bias condition of (-50 meV, 100 pA) at five representative temperatures respectively. The red triangular arrows between images show the order of measurement and all topographs are taken after annealing process. In contrast to the cold-cleaved (~10 K) images of Fig. 1d and S4, the annealing effect widens the $C_2$ orientational domain and only the $C_2$ domain aligned with the small bulk orthorhombicity persists down to 4.6 K. The inset in upper right corner in each image shows the Fourier transform, and the crosses are markers for $|\boldsymbol{q}|=2\pi/a_0$. The red arrows in inset indicate the $2a_0$ modulation ($q = \pi/a_0$), while the blue and green arrows ($q = \frac{3}{8}\frac{2\pi}{a_0}$ and $\frac{5}{8}\frac{2\pi}{a_0}$) reflect the additional ~$8a_0$ supermodulations in the $C_2$ areas appearing between $T_M$ and $T^*$. Static long-range $C_4$-symmetric (2×2) plaquette AFM order (bright areas in the bottom row **f-h**) appears below 60 K and almost completely covers the field of views below ~50 K. The insets in bottom right corner in **a-e** are representative $dI/dV$ spectra. **k**, The suggested phase diagram of $Sr_2VO_3FeAs$ based on the observation. **l-o**, The line cuts of the Fourier transforms of the temperature-dependent and tip-dependent topographs a-j, along a and b axes.

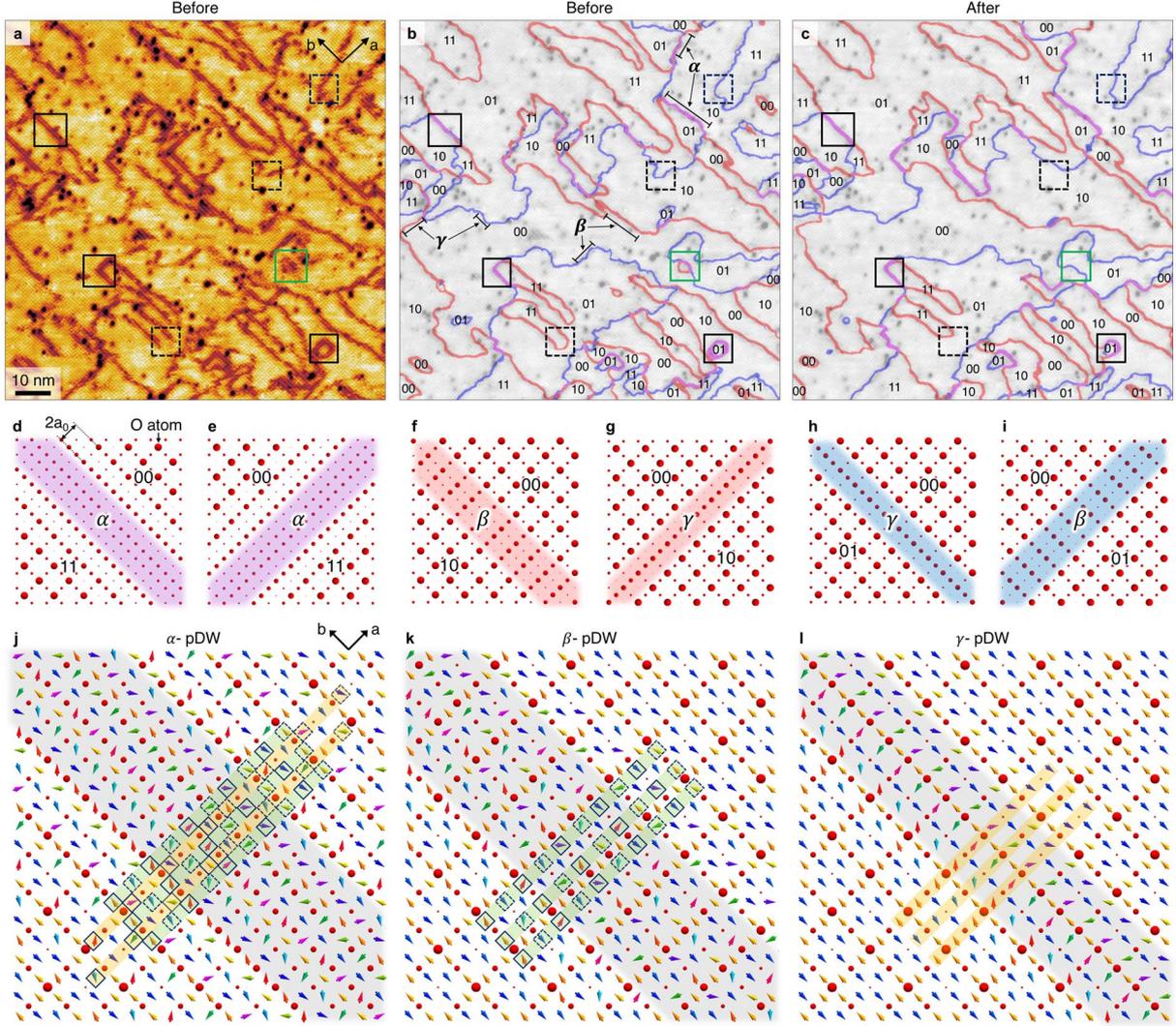

**Figure 4. Phase domain walls in plaquette AFM order. a,** SPSTM topograph taken over (106 nm)² area with bias (-50 meV, 100 pA). **b,** The pDWs detected with spatial lock-in technique for the image **a** (Supplementary Section 6). **c,** The pDWs for another SPSTM topograph taken 20 minutes later whose original topograph is in Fig. S12. The two-digit binary number $n_a n_b$ for each domain indicates the domain phase i.e. the (2×2)-unit-cell magnetic modulations shifted by $n_a$ and $n_b$ lattice unit cell in a and b directions respectively. The dashed-line (solid-line) boxes show the common areas between image **b** and **c** for mobile β- and γ-pDWs (for immobile α(purple)-pDWs). The green solid-line box shows a complex pDW change corresponding to the simulation in Fig. 5i-p. **d-i,** Relationship of two pDW classifications, one with three colors representing the relative phases of adjacent domains and the other with three Greek letters representing the symmetry of intra-pDW spin configuration. **j-l,** The low energy spin configurations for three types of straight pDWs, determined using the Landau-Lifshitz-Gilbert equation without the tip effect. The long green and yellow rectangles perpendicular to each pDW correspond to AFM and FM 1D spiral DWs respectively.

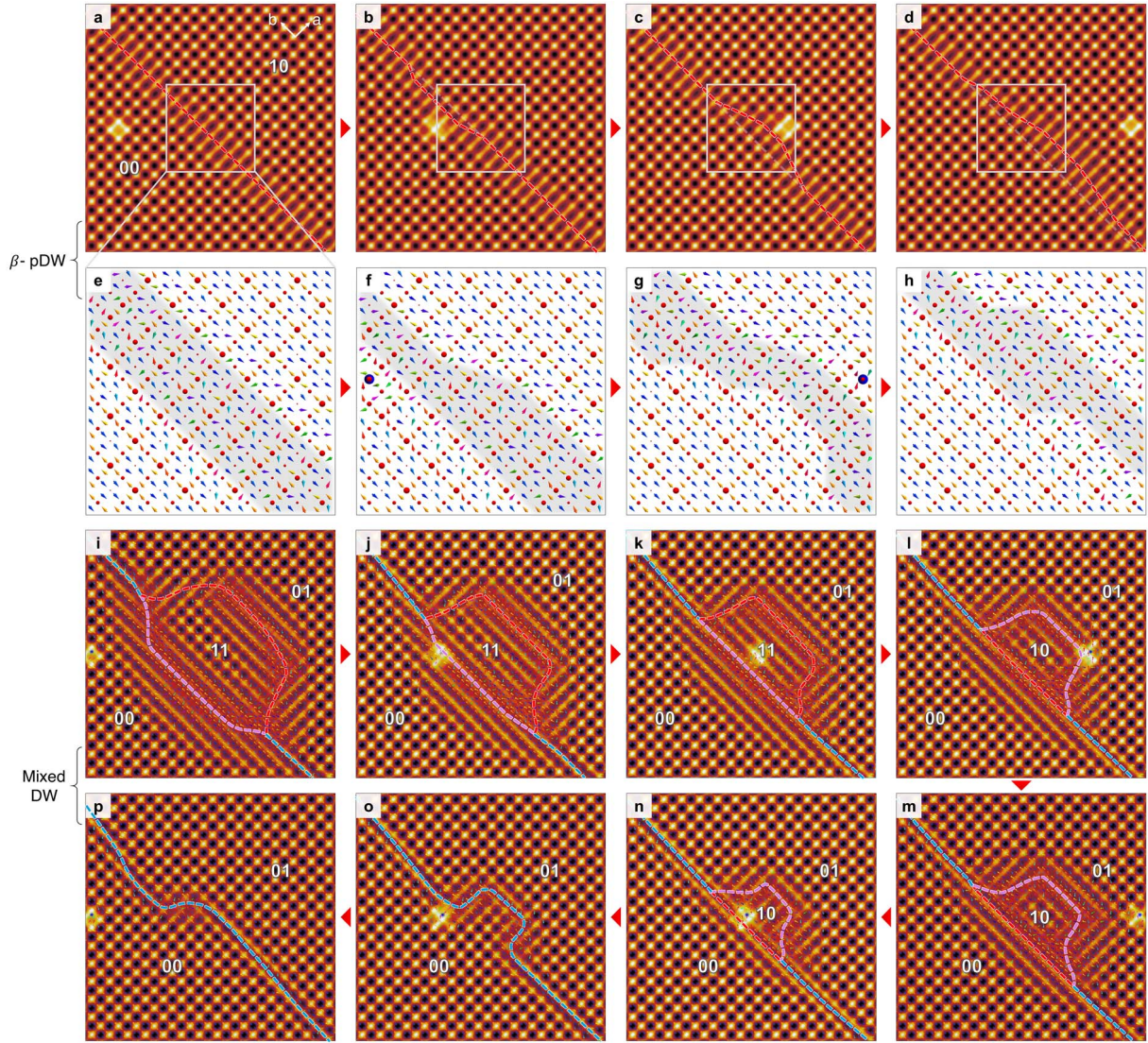

**Figure 5. Modification of pDWs induced by spin-polarized tunneling current. a-h,** A Landau-Lifshitz-Gilbert (LLG) simulation of β-pDW motion induced by the spin-torque effect of the spin-polarized tunneling current by the SPSTM tip scanned across the pDW (indicated as the horizontally moving bright extended spots in **a-d** and blue balls in **f-g**). **i-p,** A LLG simulation of erasing a localized domain island (11) on a γ-pDW by the same effect as in **a-h**, which involves multiple abrupt phase reversals (11→10→00) of the localized domain island and gives one possible explanation to the experimentally observed erasure of a localized domain island marked in the green squares in Figs. 4b-c.

**Supplementary Information**

# Atomic-scale observation and manipulation of plaquette antiferromagnetic order in iron-based superconductor

Seokhwan Choi, Won-Jun Jang, Jong Mok Ok, Hyun Woo Choi, Hyun-Jung Lee,

Se-Jong Kahng, Young Kuk, Ja-Yong Koo, SungBin Lee, Sang-Wook Cheong,

Yunkyu Bang, Jun Sung Kim, Jhinhwan Lee

## 1. Local moment based spin theories of iron-based superconductors (FeSC) and effects of quantum fluctuations and inter-sublattice interactions.

In the localized magnetic moment picture, the magnetic ground states of FeSC has been well understood by the super-exchange interactions between the Fe spins $\vec{S}_i$ and is described by $J_1$-$J_2$-$J_3$ Heisenberg model[4,6,7,19-21]

$$H = \sum_{\langle i,j \rangle} J_1 \vec{S}_i \cdot \vec{S}_j + \sum_{\langle\langle i,j \rangle\rangle} J_2 \vec{S}_i \cdot \vec{S}_j + \sum_{\langle\langle\langle i,j \rangle\rangle\rangle} J_3 \vec{S}_i \cdot \vec{S}_j \tag{1}$$

where $\langle i,j \rangle$, $\langle\langle i,j \rangle\rangle$, $\langle\langle\langle i,j \rangle\rangle\rangle$ indicate the 1st, 2nd and 3rd nearest neighbors between the two sites $i$ and $j$, and their interactions are $J_1, J_2$ and $J_3$ respectively.

Figure S1a. shows the zero-temperature phase diagram of this model in terms of $J_3/J_1$ and $J_2/J_1$. With dominant nearest-neighbor interaction $J_1$, the system prefers Néel ordering with the wavevector $\boldsymbol{Q} = (\pi,\pi)_{Fe}$ as seen in the bottom left corner of Fig. S1a. (We express the wavevectors in terms of a single Fe unit-cell basis with subscript 'Fe'. Thus, Néel order is described by the wavevector $\boldsymbol{Q} = (\pi, \pi)_{\text{Fe}}$.) With dominant second-neighbor interaction $J_2$, stripe phase is favored with ordering wavevector $\boldsymbol{Q} = (\pi, 0)_{\text{Fe}}$.

In the presence of $J_1$ or $J_2$, and finite third-neighbor interaction $J_3$ introduces frustration in the system and stabilizes two different spiral orders depending on the ratios $J_3/J_1$ and $J_2/J_1$, and they are described by the wavevectors either $\boldsymbol{Q} = (q, q)_{\text{Fe}}$ or $\boldsymbol{Q} = (\pi, q)_{\text{Fe}}$. Those four distinct phases are well understood and some of them have already been observed in FeSC materials[4,7,21]. On the other hand, in the limit of $J_1 = 0$ but finite $J_2$ and $J_3$, there are extensive degeneracies in the classical ground state with ordering wavevectors $\boldsymbol{Q} = \pm \left( \frac{\pi}{2}, \pm \frac{\pi}{2} \right)_{\text{Fe}}$. However, it turns out that quantum fluctuation lifts such accidental degeneracies of the ground states via their zero-point energies, often referred as quantum order-by-disorder mechanism[22].

As shown in the upper red area in Fig. S1a, quantum fluctuation possibly selects two collinear spin states under the conditions $J_3 > \frac{|J_2|}{2}$ and $J_3^2 \gtrsim SJ_1^2$ where $S$ is the spin magnitude; (1) diagonal double stripe (DDS) order and (2) plaquette (orthogonal double stripe, ODS) order[4]. Figure S1b,c describe such two collinear states, DDS order and plaquette (ODS) order. The DDS order is characterized by the same propagating spin wavevector, $\boldsymbol{Q} = \left( \frac{\pi}{2}, \frac{\pi}{2} \right)_{\text{Fe}}$, for both of the two Fe sublattices ($\text{Fe}_{\text{A}}$ and $\text{Fe}_{\text{B}}$) breaking $C_4$ symmetry. Whereas, the plaquette (ODS) order preserves $C_4$ symmetry and is described by two propagating spin wavevectors $\boldsymbol{Q} = \left( \frac{\pi}{2}, \frac{\pi}{2} \right)_{\text{Fe}}$ and $\left( \frac{\pi}{2}, -\frac{\pi}{2} \right)_{\text{Fe}}$ orthogonal to each other. Although both DDS order and plaquette (ODS) order are selected by quantum fluctuations, a finite but small $J_1$ causes instability in one of them[4]. In order to understand the physics of such instability, it is useful to consider their spin wave dispersions in the limit of $J_1 = 0$ first and then discuss the perturbation in the case of finite $J_1$.

Figure S1d-g show the spin wave dispersions of individual Fe sublattices ($Fe_A$ and $Fe_B$) for both DDS order and plaquette (ODS) order. There exist both protected nodes (PN's) marked with thick white circles in the direction of antiferromagnetism in the sublattice and accidental nodes (AN's) marked with thin green dashed circles in the direction of ferromagnetism in the sublattice. We note that PN's are where true Goldstone modes occur, whereas at AN's the quantum fluctuations will lift the classical zero modes. Now, we introduce a small but finite $J_1$ as a perturbation parameter. Since $J_1$ couples two different sublattices $Fe_A$ and $Fe_B$, one should consider the low energy excitations on PN's and AN's emerging at the wavevectors $Q = \pm\left(\frac{\pi}{2}, \frac{\pi}{2}\right)_{Fe}$ or $\pm\left(\frac{\pi}{2}, -\frac{\pi}{2}\right)_{Fe}$. For plaquette (ODS) order, every PN couples to an AN, thus the perturbation with $J_1$ remains non-singular because the zero modes in AN are always lifted by quantum fluctuations. This allows the plaquette (ODS) order as a stable state in a finite range of $J_1$. However, the perturbation with $J_1$ makes the DDS order unstable since true Goldstone modes at PN's of both sublattices are coupled, introducing singular perturbation on degenerate spin wave states. The result is the unstable Goldstone modes with complex energy which makes the DDS order itself unstable. Hence, the plaquette (ODS) order is uniquely selected in the presence of finite $J_1$ and quantum fluctuations. We note that neither further neighbor interactions ($J_i$, $i > 3$) nor orthorhombic structural distortions can naturally select the plaquette (ODS) order. In detail, the further neighbor interactions leave the ODS and DDS states still degenerate, and an orthorhombicity which allows $J_{2a} \neq J_{2b}$ (See Fig. S1c) selects the DDS order with energy gain proportional to $-|J_{2a} - J_{2b}|$ by choosing its antiferromagnetic spin wavevector common to both sublattices in the direction of larger $J_2$.

As mentioned above, orthorhombicity with anisotropic $J_2$ (which breaks $C_4$ symmetry) favors the DDS order to minimize spin-spin interactions. Indeed, such orthorhombic distortions are present in most FeSC below ~150 K and they strongly influence the symmetry of the spin structures at low temperatures[6,7,23]. It is mainly due to this reason that, surprisingly, the plaquette (ODS) order has never been observed explicitly in FeSC, except for some indirect signatures found in neutron scattering experiments[36]. Thus, the absence of strong orthorhombic distortion below magnetic ordering temperature may open a possibility to find the $C_4$ symmetric plaquette (ODS) order as a strong evidence for the presence of quantum effect. Furthermore, it may play a key role in the study of the interplay between magnetic order, superconductivity and their quantum fluctuations in the absence of magnetoelastic effect associated with structural distortion.

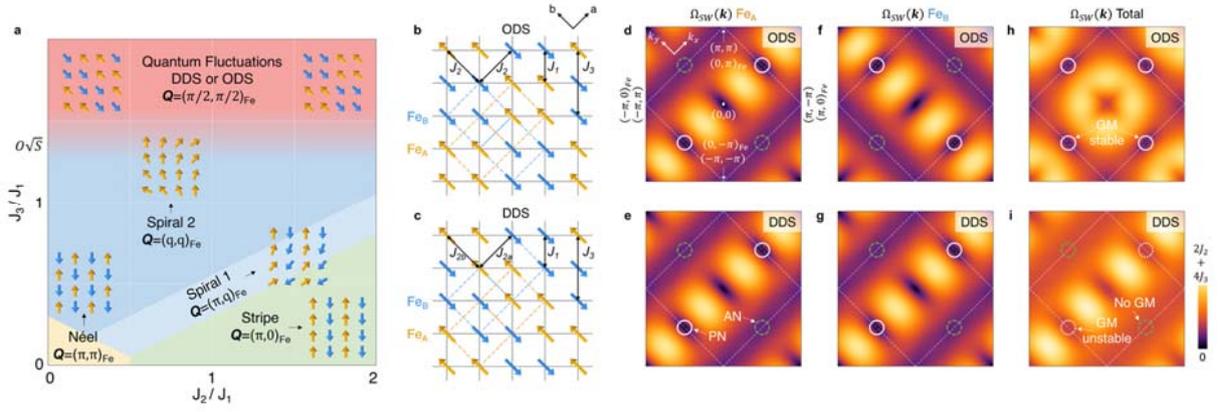

**Figure S1. a**, Zero-temperature phase diagram of $J_1$-$J_2$-$J_3$ Heisenberg model. In the limit of dominant $J_3$, quantum fluctuations select two collinear states; plaquette (ODS) order and diagonal double stripe (DDS) order. **b,** Schematic description of plaquette order where spin wavevectors in $Fe_A$ and $Fe_B$ are orthogonal with each other. Atomic (1×1) and magnetic (2×1) unit cells are marked with blue ($Fe_A$) and red ($Fe_B$) dashed boxes. **c**, Schematic description of DDS order where spin wavevectors for both sublattices are the same. Distinction between $J_{2a}$ and $J_{2b}$ are introduced for the argument related to orthorhombic structural distortion. **d-g**, Spin wave dispersion $\Omega_{SW}(\boldsymbol{k})$ of the two sublattices for plaquette (ODS) order and DDS order in the limit of $J_1$=0. There exist the protected nodes marked as PN (white solid circle) and the accidental nodes marked as AN (green dashed circle). PN is related to the true Goldstone modes, whereas the classical zero modes at AN will be lifted by quantum fluctuations. **h-i**, Spin wave dispersion $\Omega_{SW}(\boldsymbol{k})$ combining both contributions of $Fe_A$ and $Fe_B$ for plaquette (ODS) order and DDS order. For the plaquette (ODS) order, coupling between PN of $Fe_A$ and AN of $Fe_B$ (or vice versa) via $J_1$ leaves the Goldstone modes of PN stable, making the system remain stable. However, in case of DDS order, the PNs of $Fe_A$ and $Fe_B$ are coupled via $J_1$ with degenerate nodal dispersions. This makes the perturbation singular and leads to instability of the Goldstone modes in the DDS order.

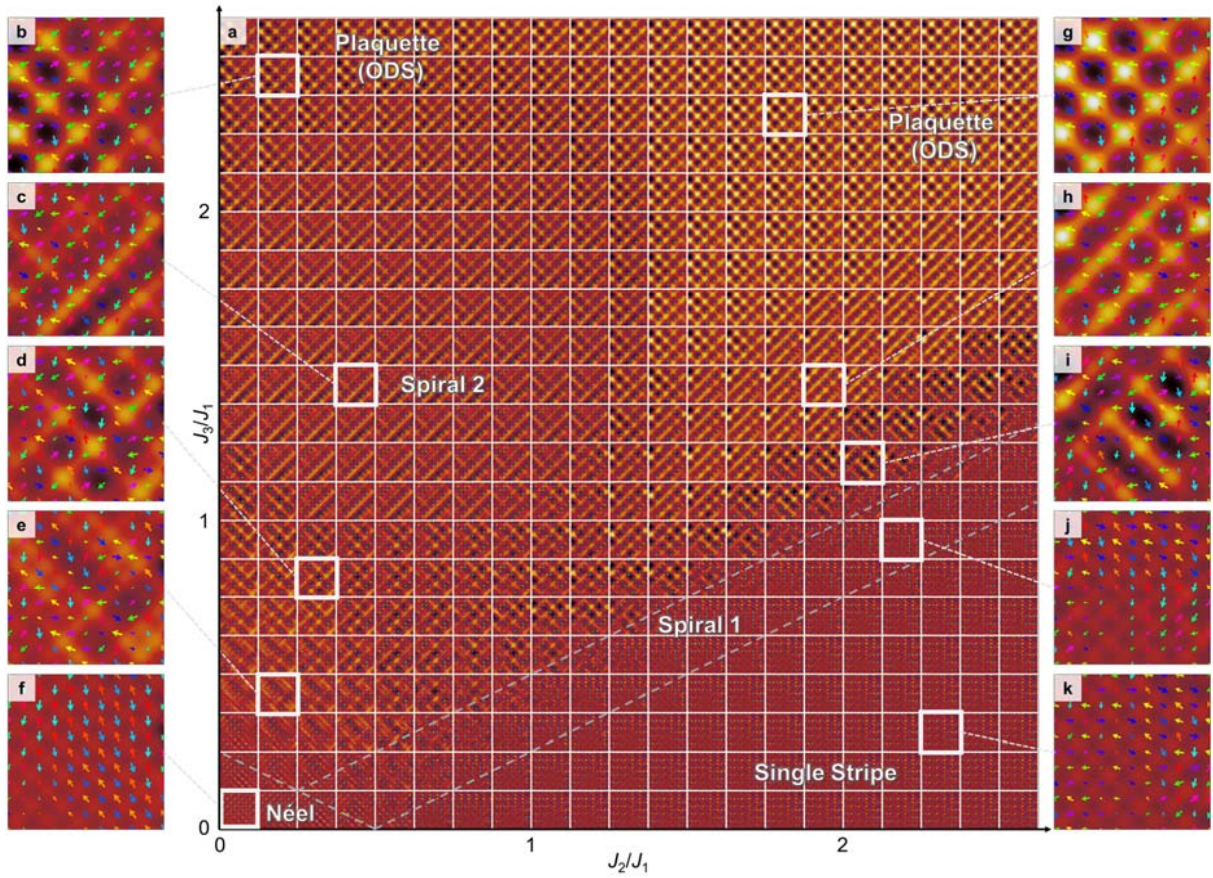

**Figure S2. a,** A simulated phase diagram of $J_1$-$J_2$-$J_3$ Heisenberg model using the same code as for the Figs 4-5 and Figs S14-S15. Each small square image shows a low energy spin configuration relaxed from random initial and boundary conditions for ($J_2/J_1$, $J_3/J_1$) defined at its lower left corner. **b-k,** Magnified views of the spin configurations indicated by white squares in **a**. In the upper part of this phase diagram, ODS, DDS and spiral 2 orders are nearly degenerate in energy and the actual spin orders obtained by LLG relaxation are slightly dependent on random initial and boundary spin configurations used.

## 2. Preparation of antiferromagnetic (AFM) Cr tip for SPSTM

There have been several reports[23,37] on measurement of AFM order and its domain walls with a Fe-cluster-terminated tip whose spin polarization is controlled with applied magnetic field. However, the stray magnetic field due to the ferromagnetic Fe film or cluster may disturb the sample and interfere with the accurate measurement of the true magnetic ground states. In particular, it leaves controversial issues in determining the magnetic ground state for the systems where multiple magnetic orders and/or type II superconductivity compete with one another, since vortices can be always formed close to the ferromagnetic Fe tip.

In order to overcome such problems, we have developed AFM Cr-cluster-terminated tip which has negligible stray field but is more challenging to achieve spin polarization. The negligible stray field with Cr-cluster tip is a big advantage in accurate SPSTM measurements since such AFM tip can capture the delicate spin states close to the domain walls.

The spin-polarized Cr cluster tip was prepared by collecting Cr atoms on the apex of the W tip by controlled field emission with parameters depending on the sharpness of the base W tip. The Cr cluster tip was then tested for proper in-plane spin-polarization by observing multiple levels of differential conductance at set point bias near -50 meV on multiple antiferromagnetic terraces with identical orientation on a stepped Cr surface as shown in Fig. S3.

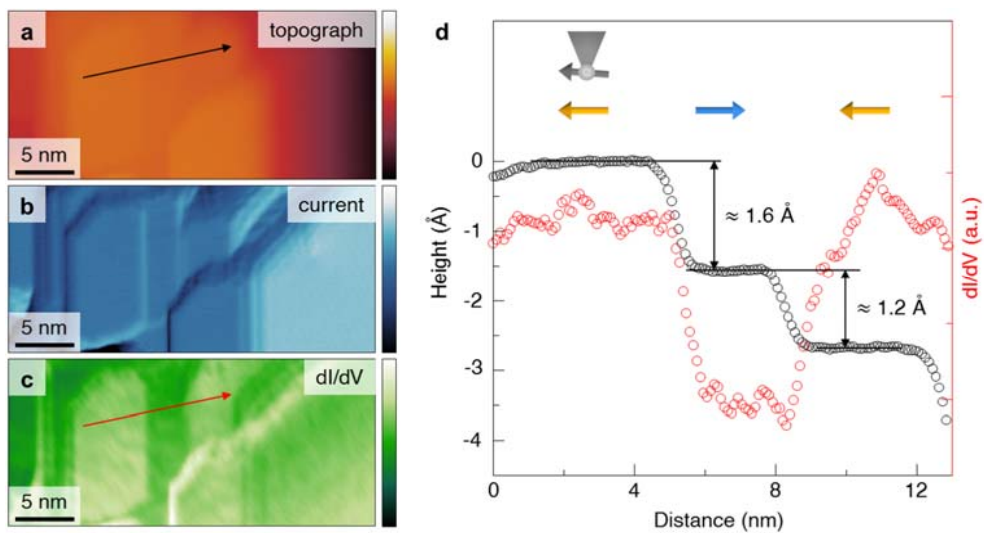

**Figure S3.** A sputter-annealed Cr surface topograph, **a**, and simultaneously taken current image, **b**, and dI/dV image, **c**, at bias condition of (-50 meV, 200 pA). **d**, The spin-contrast shown in cross-sections of topograph and dI/dV along a line marked with arrows in **a** and **c**.

### 3. Low temperature cleaving of Sr₂VO₃FeAs

The $Sr_2VO_3FeAs$ sample with size about 250 μm × 250 μm was glued on the sample holder using silver epoxy and a cleaving rod was glued on top of it using identical epoxy. The sample holder was cooled down to ~ 10 K by contacting a copper block at 4.2 K for ~ 2 hour. The cleaved surface was qualitatively identical as shown in Fig.1d for all 15 cleaves. It means that the surface shown in Fig. S4 is the SrO terminated surface which is the only symmetric and charge-neutral cleavage surface in $Sr_2VO_3FeAs$.

In the case of the $BaFe_2As_2$-based FeSCs[38], there is no charge-neural symmetric cleavage plane and more than one kinds of surface reconstruction can be seen as a result. However, in $Sr_2VO_3FeAs$, it is always the SrO-SrO bilayer that is cleaved in half, similarly to the case of BSCCO where the BiO-BiO bilayer is always cleaved resulting in charge neutral cleaved surface. From the cleavage statistics (Fig.S4) and the tip dependent and the bias dependent SR changes mentioned Fig.2, it is clear that the different surface structures in Fig. 1d and 1e are not from different cleaved surfaces but due to different tip spin polarization conditions.

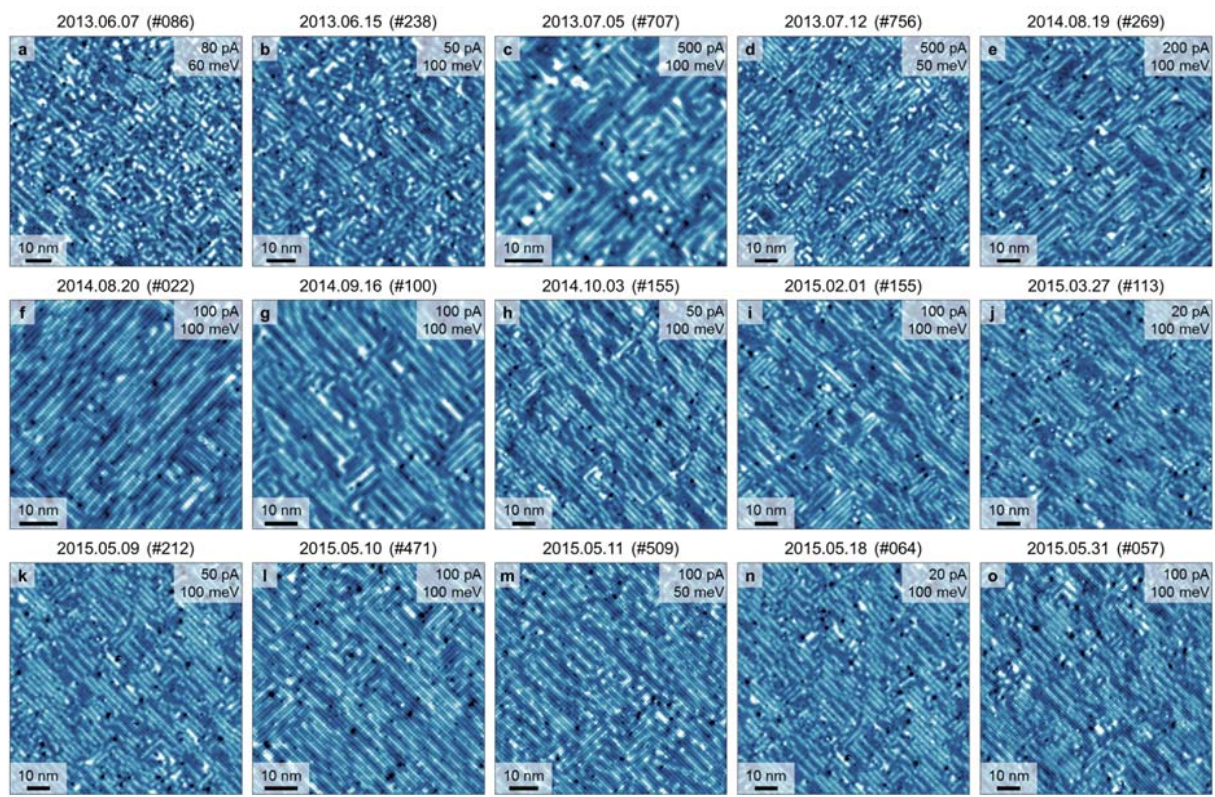

**Figure S4.** All 15 cleaved surfaces of 6 single crystals showed virtually identical topographs under W tip, confirming the scenario of symmetrical cleavage at the SrO-SrO bilayer. The chance of observing 15 identical surfaces when there is 50:50 chance of observing one of two different asymmetrically cleaved surface is less than 0.00611% or more exactly, 1 in 16,384.

## 4. Characteristics of plaquette (ODS) order observed in $Sr_2VO_3FeAs$

In the Cr-cluster tip SPSTM imaging, the vertical tunneling conduction channel made of O-V-As to the Fe layer equally samples the spin polarization of the four Fe atoms neighboring of each As atom. When every four Fe spins underneath a particular O-V-As chain are parallel (antiparallel) to the Cr tip's spin polarization, the top layer O atom will look bright (dark) in the SPSTM topograph. On the other hand, if the four Fe spins underneath an O-V-As chain are grouped into two roughly parallel spins and two roughly antiparallel spins, they will possess neutral brightness just at the average of the brightness of the bright and the dark O atoms.

In this condition, the plaquette (ODS) order in Fe layer generates a ($2\times2$) magnetic unit cell with characteristic intra unit cell pattern as simulated and shown in Fig. S7a. In contrast, the DDS (diagonal double stripe) order, the Néel order, the single stripe order, and the two kinds of spiral orders that may appear in the spin-order phase diagram for Heisenberg model have qualitatively different magnetic unit cells as shown in Fig. S7b-f.

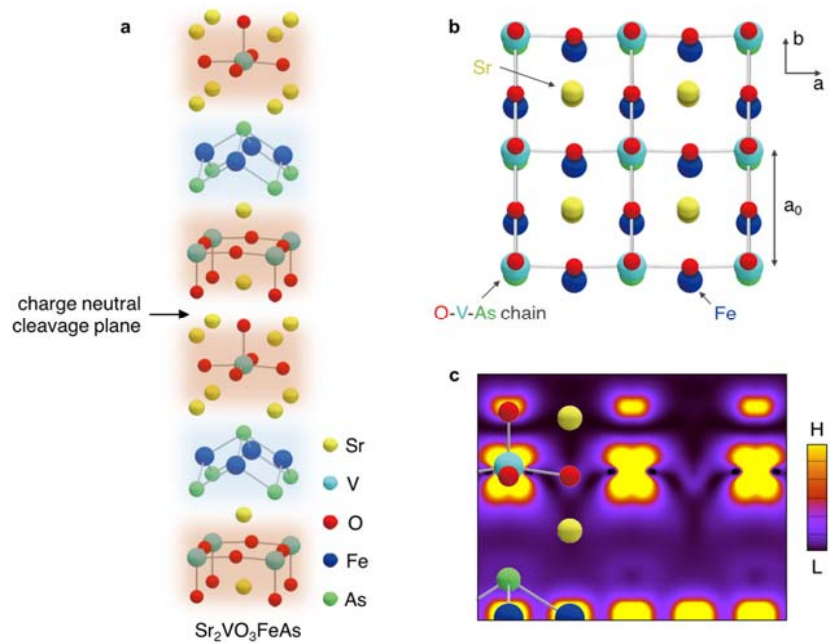

**Figure S5. a,** The atomic lattice structure of $Sr_2VO_3FeAs$. It has a charge neutral cleavage plane between the two SrO layers in contact. **b**, Tilted top-view of $Sr_2VO_3FeAs$. **c**, The LDA-based electron density plot near the Fermi level (integrated over [-50,0] meV), showing the existence of the vertical tunneling path made of O (top SrO layer) – V ($VO_2$ layer) – As (FeAs layer).

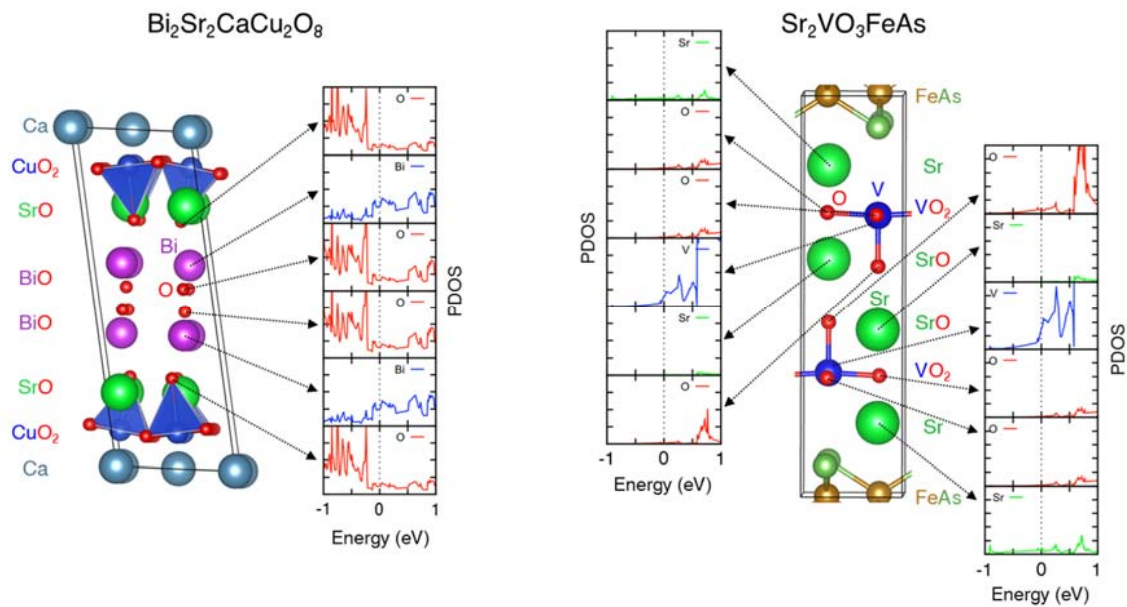

**Figure S6.** LDA-based projected density of states (PDOS) for (**a**) $Bi_2Sr_2CaCu_2O_8$ (BSCCO-2212) and (**b**) $Sr_2VO_3FeAs$. $Bi_2Sr_2CaCu_2O_8$ has higher PDOS on Bi compared with O on the top cleaved surface, while $Sr_2VO_3FeAs$ has higher PDOS on O compared with Sr on the top cleaved surface.

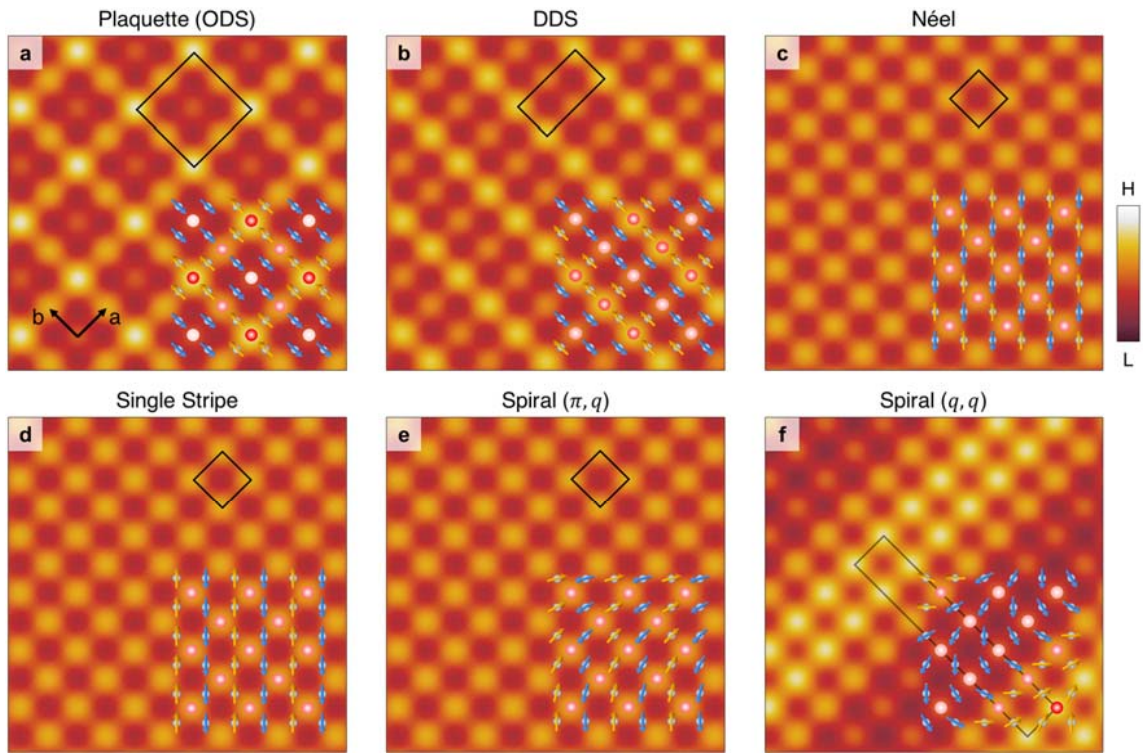

**Figure S7.** Simulated SPSTM topographs for various possible magnetic orders in the tunneling geometry of Sr$_2$VO$_3$FeAs. **a**, Plaquette (orthogonal double stripe, ODS) order, **b**, DDS (diagonal double stripe) order, **c**, Néel order, **d**, Single stripe order, **e**, Spiral order (I) with wavevector $(\pi, q)$, **f**, Spiral order (II) with wavevector $(q, q)$.

**5. Conversion of heavily surface reconstructed surface into fluctuating, effectively flat surface by spin-polarized tunneling current**

Reflecting the strong coupling of the surface reconstruction (SR) with the spin degrees of freedom inside the perovskite layer that lies between the tip and the FeSC layer, we observed significantly different bias thresholds for SR fluctuations for tips with and without spin-polarization. With the W tip, we observed that the SR starts to fluctuate with a bias threshold $V_W \sim 280$ mV as shown in Fig. S8 below. The SR fluctuation leads to disordered SR as shown in Fig. S8i. On the other hand, with the Cr tip, we observed the SR equivalent to that shown with W tip below $V_W$ only when the bias is below $V_{Cr1} \sim 40$ mV as shown in Fig. S9 below. Above $V_{Cr1}$, the surface changes to an apparently flat uniform SR with (2×2) domains and phase domain walls visible. As we go back to bias much smaller than $V_{Cr1}$, the surface keeps the same topographic characteristics except that we can see individual SR fluctuation as horizontal streaks as shown in S8i. This implies that the apparently flat uniform SR appearing with the (2×2) domains and pDWs at bias larger than $V_{Cr1}$ in case of Cr tip is due to lowered threshold ($V_{Cr1} \rightarrow V_{Cr2}$, $V_{Cr2} < 10$ mV) as the bias is increased beyond the initial $V_{Cr1}$, which then causes rapid SR fluctuations for all bias larger than $V_{Cr1}$ since $V \gg V_{Cr2}$ is satisfied for $V \gg V_{Cr1}$. As can be seen in Fig. S10, the tips (W and Cr) show no change in their characteristics before and after the SR modifications at high bias.

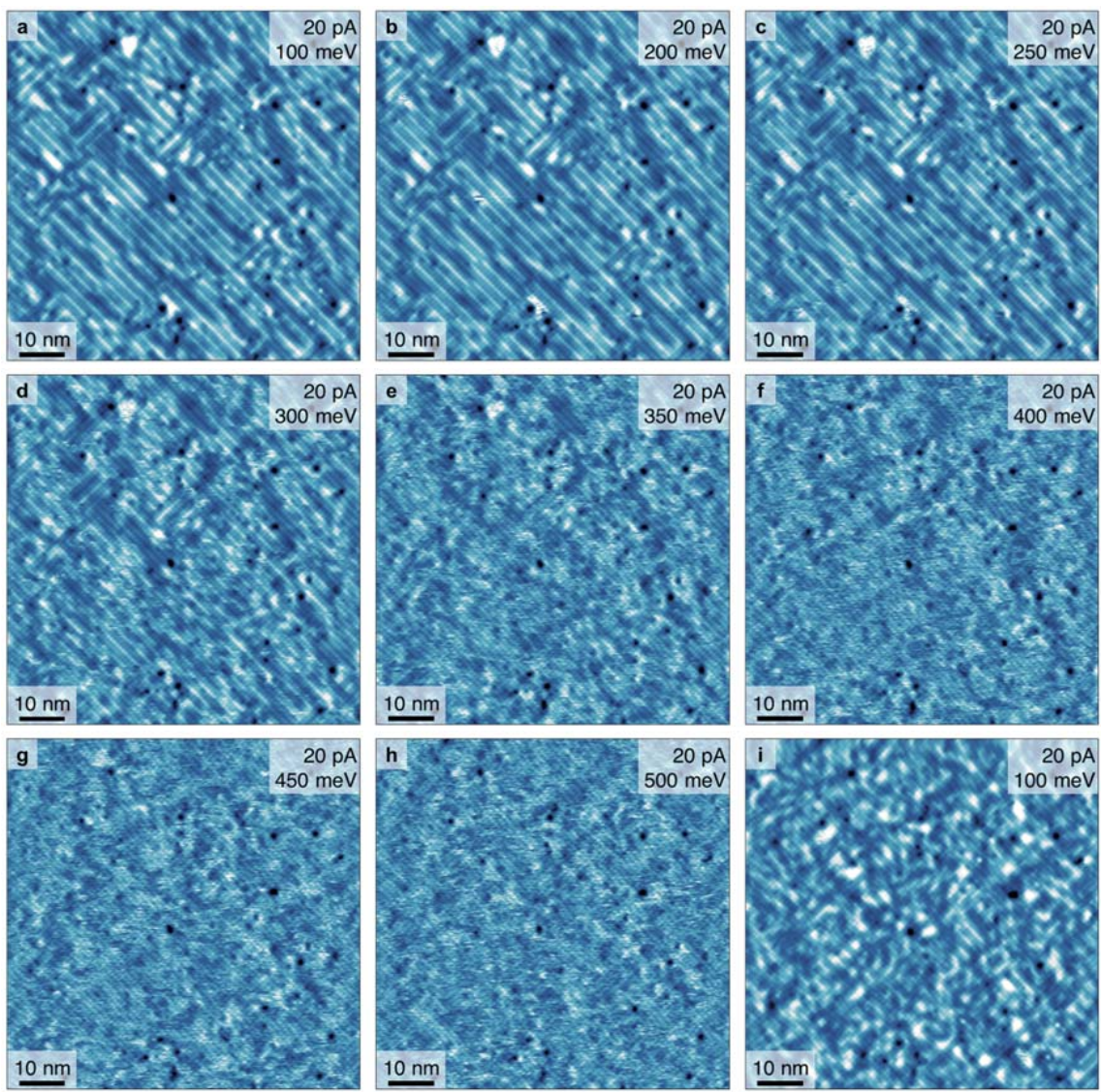

**Figure S8.** W-tip based experiment on evolution of surface topograph over an identical area with varied bias conditions. The images are listed in the order of measurement. Only above 280 meV bias voltage the surface start to change significantly. Above 400 meV bias voltage the surface fluctuates significantly, making it appear virtually flat and free from SR. However, when the bias voltage is reduced below the 280 meV threshold again, the surface is frozen in a state with randomized SR.

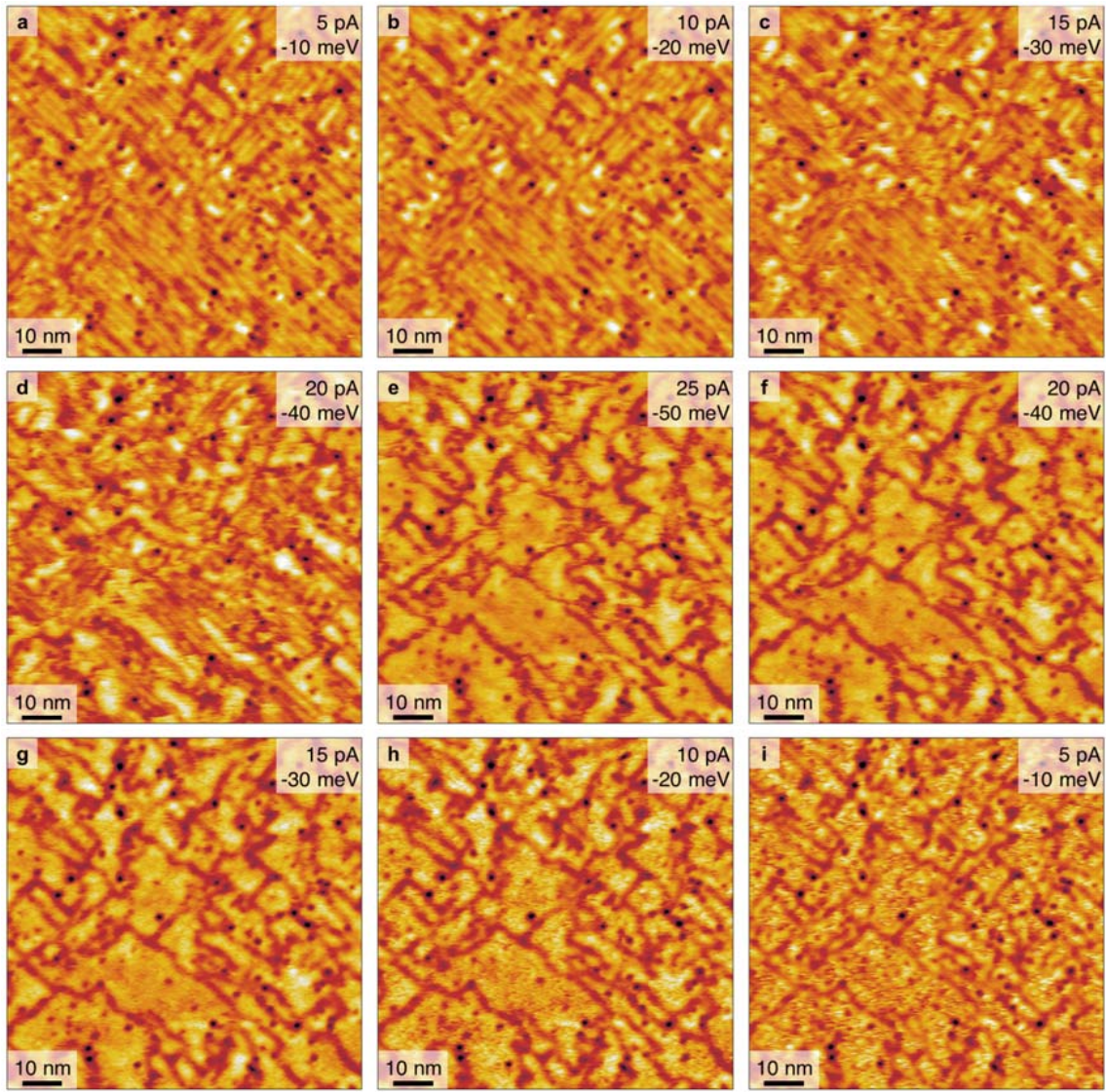

**Figure S9.** Cr-tip based experiment on evolution of surface topograph over an identical area with varied bias conditions. The images are listed in the order of measurement. Just above 30 meV bias voltage the surface start to change significantly. Above 50 meV bias voltage the surface fluctuates significantly, which means that the fluctuating SRs averages out into an effectively flat surface and no longer add as a static (2×n) modulation to the Cr tip topograph in this bias range. This makes only the spin-polarized tunneling contrast to show clearly in the absence of the original SR modulations. The fluctuation of the SR slows down but still clearly visible as horizontal streaks when the bias voltage and current become weak again as shown in the last image. This implies that the surface modified by the spin-polarized current of medium energy (>40 meV) in case of Cr tip has significantly lower bias threshold for the surface fluctuation compared with the original pristine surface. At any bias voltage above this new lower threshold level, the fluctuation of the SR averages out to a uniform value over the whole surface, making only the underlying (2×2) magnetic contrast to appear significantly.

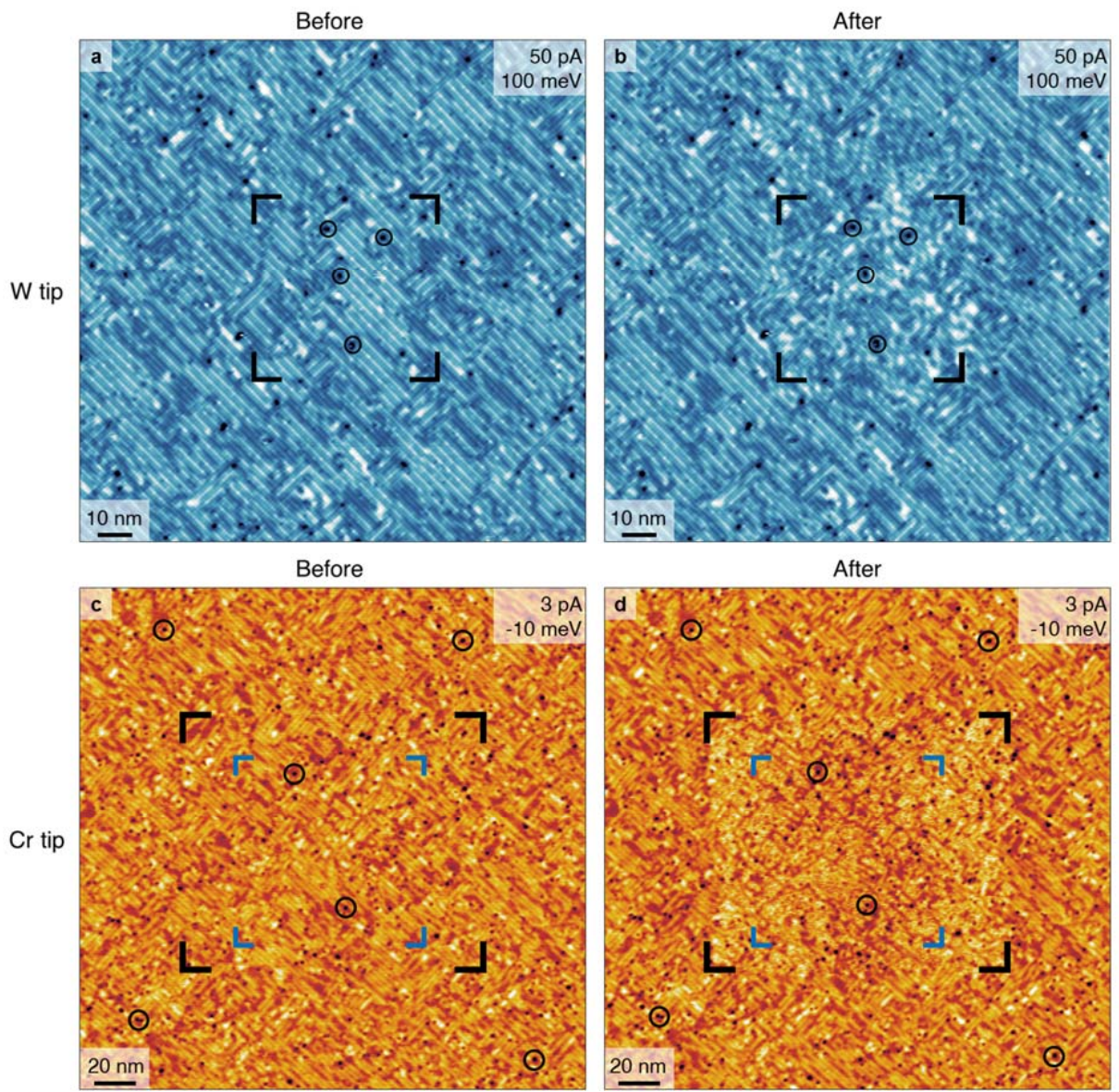

**Figure S10.** Wide area images showing the changes in the surface topographs as a result of taking the high bias images near the center. The black elbows in all four images mark the boundaries of the high bias scanning with W and Cr tips respectively. In the Cr tip case, the blue corner markers correspond to the image crop areas for Fig. S9 while the above-threshold bias imaging was performed over slightly wider area shown with the black corner markers. In both cases the areas outside the high bias scanned areas are unchanged and implies that there has been no tip changes.

## 6. Domain wall detection and visualization by spatial lock-in technique

The domain walls of the plaquette (ODS) order can be determined by first detecting the (2×2) spatial modulation phases ($\phi_a(\vec{r}), \phi_b(\vec{r})$) with a technique similar to the time-domain phase detection method known as the lock-in technique. To detect the modulation phase $\phi_a(\vec{r})$ in $a$-direction, we choose the Bragg peak position $\boldsymbol{q}_{(\pi,0)}$ corresponding to $(\pi, 0)$ in the Fourier Transform (FT) of the SPSTM topograph. We then generate two arrays $S$ and $C$ with identical size as the original topograph $T$, where $S$ and $C$ are filled with $\sin(\boldsymbol{q}_{(\pi,0)} \cdot \boldsymbol{r})$ and $\cos(\boldsymbol{q}_{(\pi,0)} \cdot \boldsymbol{r})$ respectively. Pixel-by-pixel multiplication of $T$ and $C$ denoted as $C_T$ (and $T$ and $S$ denoted as $S_T$) contains fast spatial modulations with modulation wave vector near $2\boldsymbol{q}_{(\pi,0)}$ and slow spatial modulations with wave vector near 0. As with the time-domain lock-in technique, we filter out the fast $2\boldsymbol{q}_{(\pi,0)}$ modulations with a spatial low pass filter with cut off wave vector $\sim\boldsymbol{q}_{(\pi,0)}$ and denote them $\langle C_T \rangle$ and $\langle S_T \rangle$. Such $\langle C_T \rangle$ and $\langle S_T \rangle$ contain information of $A\cos\phi_a(\vec{r})$ and $A\sin\phi_a(\vec{r})$ respectively and the phase $\phi_a(\vec{r}) = \tan^{-1}(\langle C_T \rangle, \langle S_T \rangle)$ can be defined at every pixel of the topograph as shown in Fig. S11c. The domain walls (red curves in Fig. S11b) determined by the spatial modulation phase shift in $a$-direction can then be defined as the collection of pixels with abrupt reversal (change by $\sim\pi$) of phase $\phi_a(\vec{r})$ within a magnetic unit cell distance from the pixel. Applying the identical method starting with the Bragg peak position $\boldsymbol{q}_{(0,\pi)}$ will generate $\phi_b(\vec{r})$ (Fig. S11d) and the blue domain walls (blue curves in Fig. S11b). The purple domain wall results from overlapping red and blue domain walls.

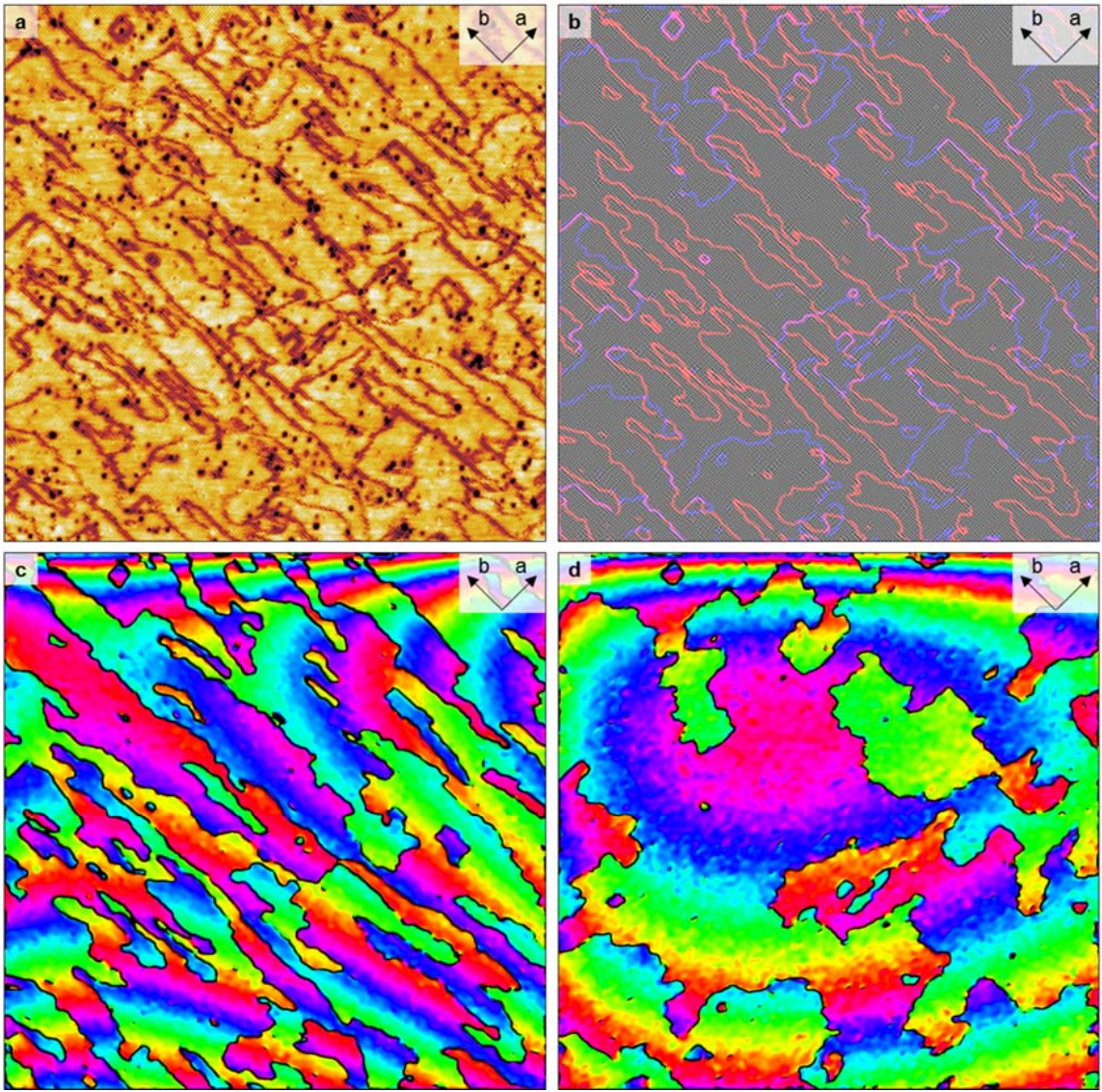

**Figure S11. a**, Large area (142 nm x 142 nm) topograph taken at 4.6 K with spin-polarized Cr tip at bias condition of (-50 meV, 100 pA). **b**, image of the automatically detected domain walls showing the (2×2) atomic modulations in each domain. **c,d**, The spatial modulation phase maps of $\phi_a(\vec{r})$ (**c**) and $\phi_b(\vec{r})$ (**d**), used to automatically detect the two types of domain walls. The piezo-creep-induced lattice distortion induces slow variations of the phases over the field of view, which do not affect the domain wall detection algorithm relying on the abrupt phase change by $\pi$ within the width (~$2a_0$) of the domain walls.

## 7. Observation of phase DW motions induced by scanned spin-polarized current

As we repeat scanning over the same area with Cr tip, some portions of the phase domain walls (pDWs) of plaquette order were subject to shifts and mergers in almost random fashions. This phenomenon can be understood by the meta-stability of the Fe spin configurations near the pDWs due to competition between two neighboring domains: Fe spins near the pDW can conform to either phase domain with very small energy difference although there will typically be a certain energy barrier. Using the spatial lock-in technique explained in Section 6, we could visualize the dynamics of the pDW motions clearly as shown in Figs. S12 and S13.

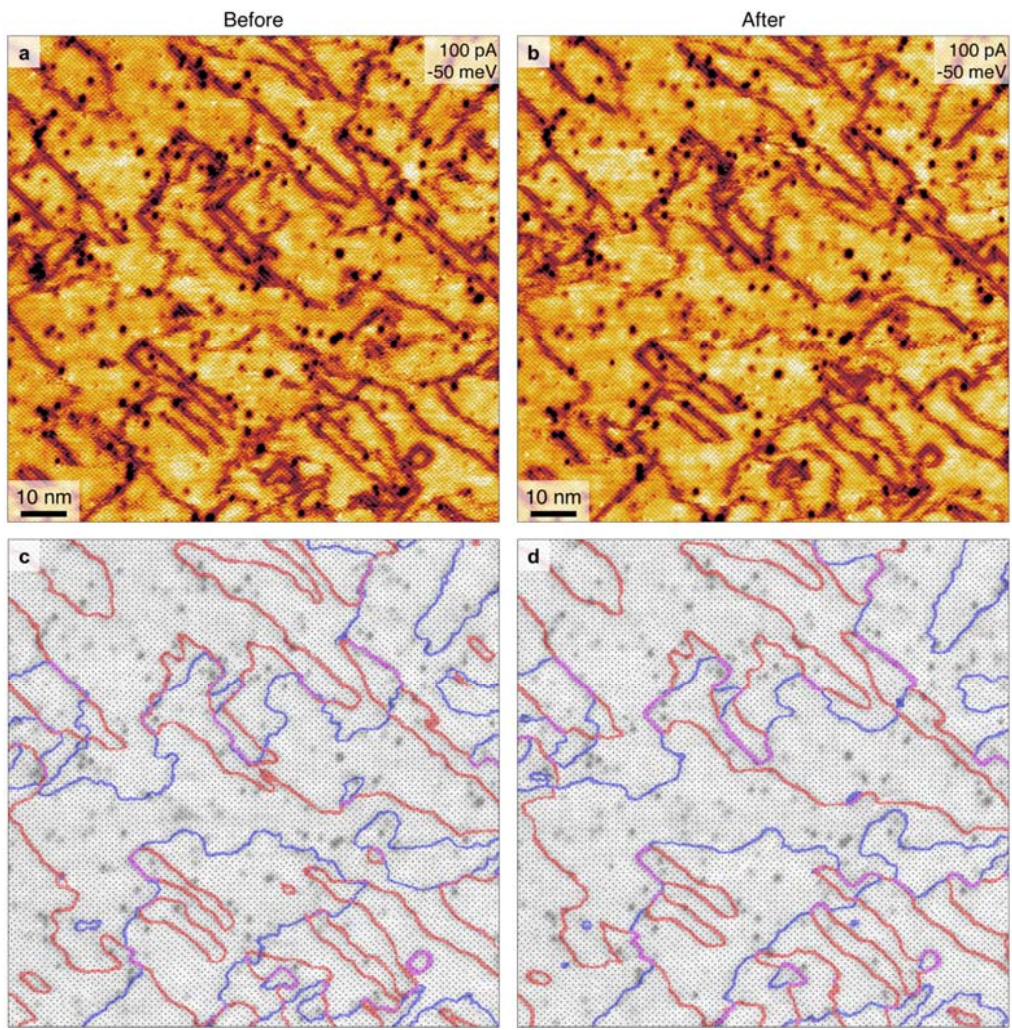

**Figure S12. a,b,** Raw data of Cr-tip SPSTM topographs used in Fig. 4b-c, taken 20 minutes apart. **c,d,** The pDWs found and colored by the domain wall detection algorithm explained in Section 6.

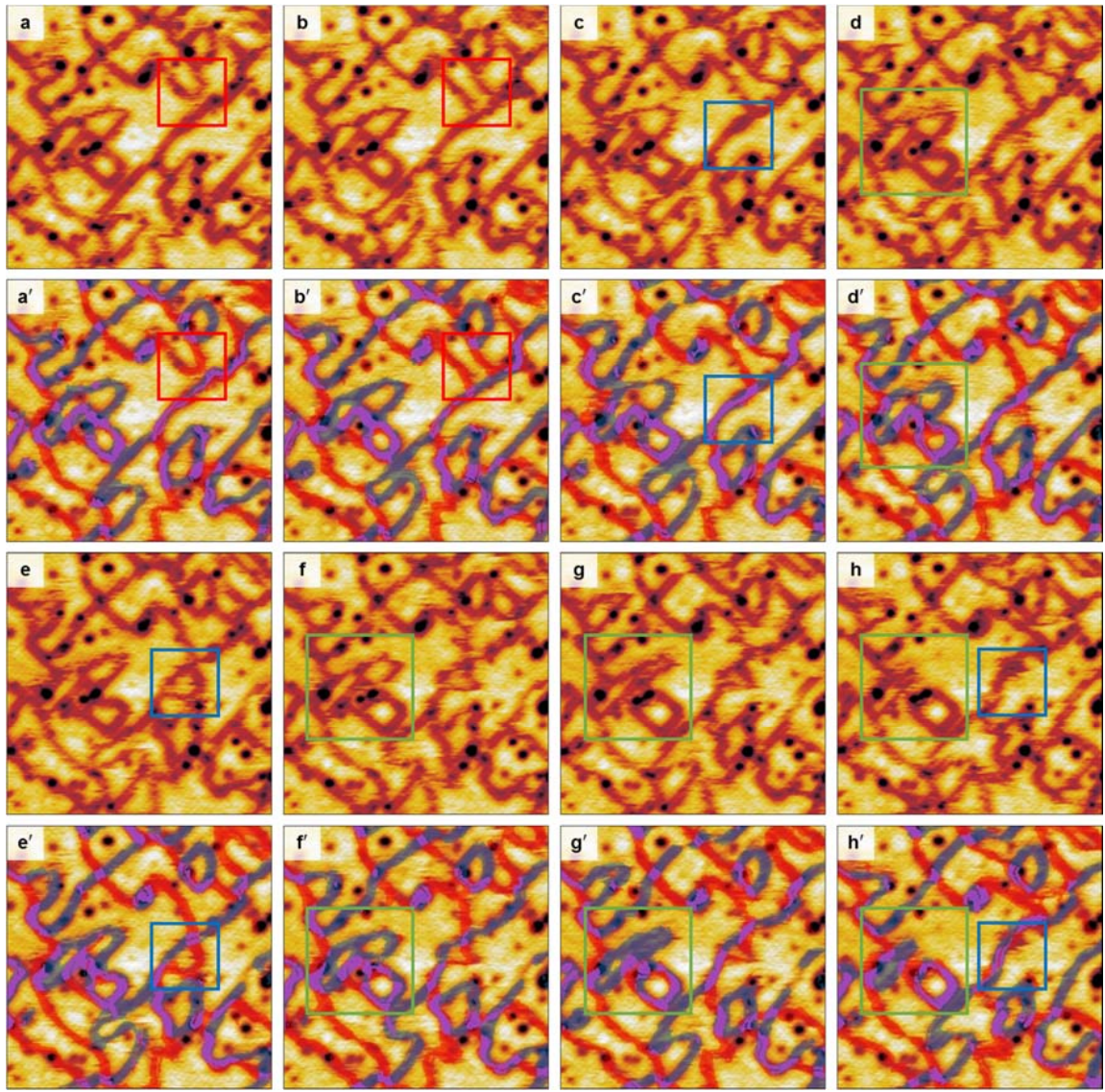

**Figure S13. a-h,** Raw data of Cr-tip SPSTM topographs taken 21 minutes apart with bias condition [-50 meV, 20 pA]. The colored squares are guides to the eyes for the topological variations of the pDWs. **a′-h′,** Types of pDWs found by the domain wall detection algorithm explained in Section 6.

## 8. Simulation of phase DW motions induced by scanned spin-polarized current

The Hamiltonian we used for the simulation is of the form

$$\mathcal{H} = J_1 \sum_{\langle i,j \rangle} \vec{S}_i \cdot \vec{S}_j + J_2 \sum_{\langle\langle i,j \rangle\rangle} \vec{S}_i \cdot \vec{S}_j + J_3 \sum_{\langle\langle\langle i,j \rangle\rangle\rangle} \vec{S}_i \cdot \vec{S}_j - K \sum_i \left( \vec{S}_i \cdot \vec{S}_0 \right)^2$$

where the last term emulates the preference of collinearity of all spins due to order by disorder effect and $\vec{S}_0$ is a mean-field collinear spin orientation for the given field of view.

For the spin simulation we used an extended Landau-Lifshitz-Gilbert (LLG) equation of the form[35]

$$\frac{\partial \vec{S}_i}{\partial t} = -\frac{\gamma}{(1+\alpha^2)\mu_S} \vec{S}_i \times \left[ \vec{H}_i + \alpha \left( \vec{S}_i \times (\vec{H}_i + \vec{T}_i) \right) \right]$$

where $\gamma$ is the gyromagnetic ratio, $\alpha = 0.025$ is the Gilbert damping constant and $\vec{H}_i = -\partial\mathcal{H}/\partial\vec{S}_i$ is the local field for spin $\vec{S}_i$ by the neighboring exchange-interacting spins. Here $\vec{T}_i$ is the effective local field representing the spin-torque effect by the spin-polarized current defined by

$$\vec{T}_i = -T_0 e^{-2\kappa'\sqrt{(x_i - x_{tip})^2 + (y_i - y_{tip})^2 + h^2}} \hat{P}$$

where $T_0$ is the effective spin-torque field strength proportional to both the spin-polarized tunneling current and the magnitude of its spin-polarization. $\hat{P}$ is the spin polarization orientation and $(x_{tip}, y_{tip})$ is the tip position. The effective decay constant $\kappa'$ and the effective tip height $h$ determine the range of spin-torque effect.

We used typical ratios of $J_2/J_1$=1, $J_3/J_1$=1.5 and $K/J_1$=0.5 for all the simulations in this paper (Figs 4-5 and S14-S15), except that $K$ was set to zero for the phase diagram in Fig. S2 for comparison with well-known classical phase diagram. We also assumed that the spin-torque effect range is a few Fe lattice constants and its strength is a fraction of the local spin field strength to simulate the domain wall motion observed by the experiment while not saturating the Fe spins under the tip.

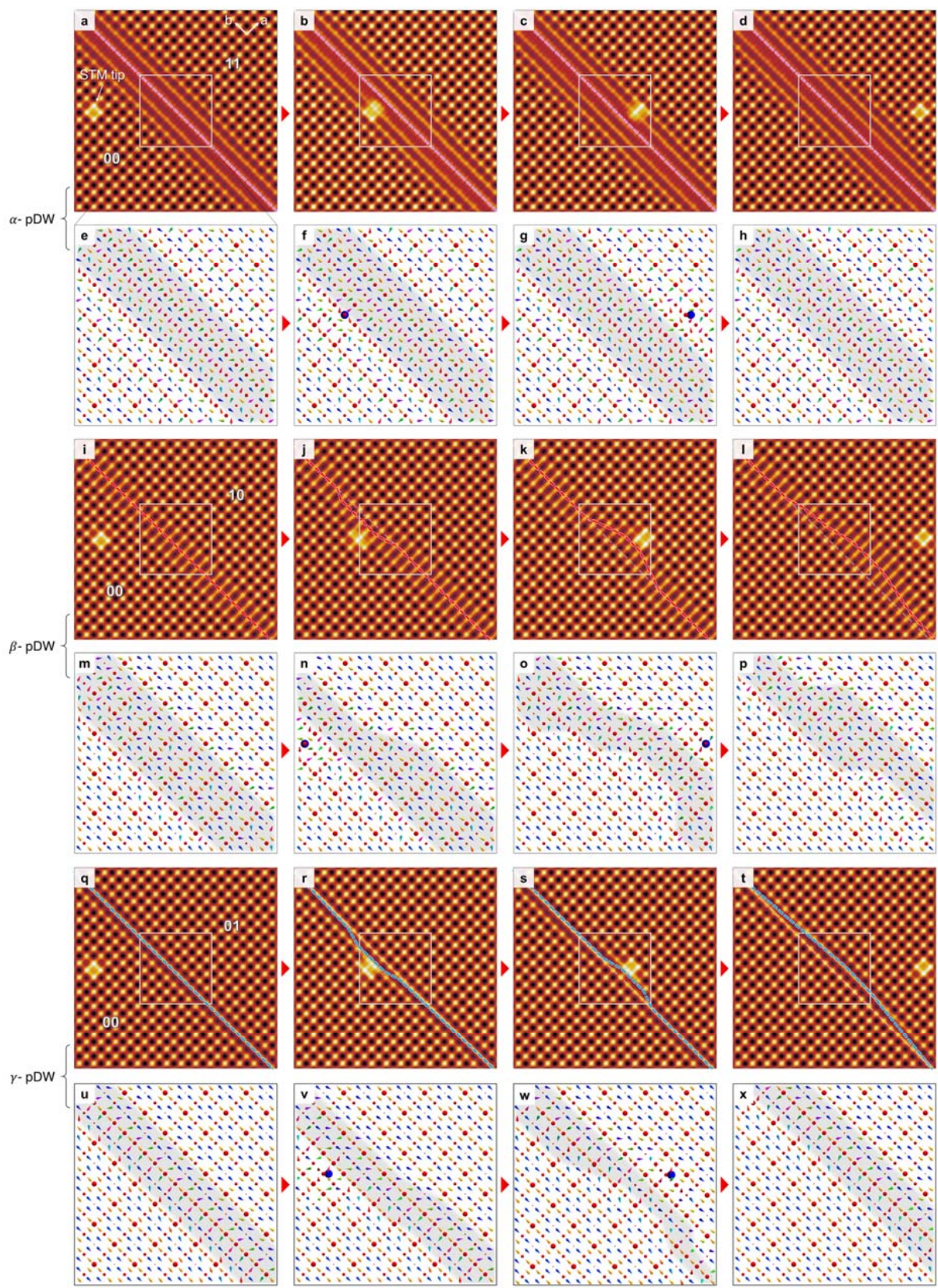

**Figure S14.** Simulated motion of various pDWs induced by scanned spin-polarized tunneling current. α-pDW (**a-h**) shows least chance of induced shift compared with β-pDW (**i-p**) or γ-pDW (**q-x**).

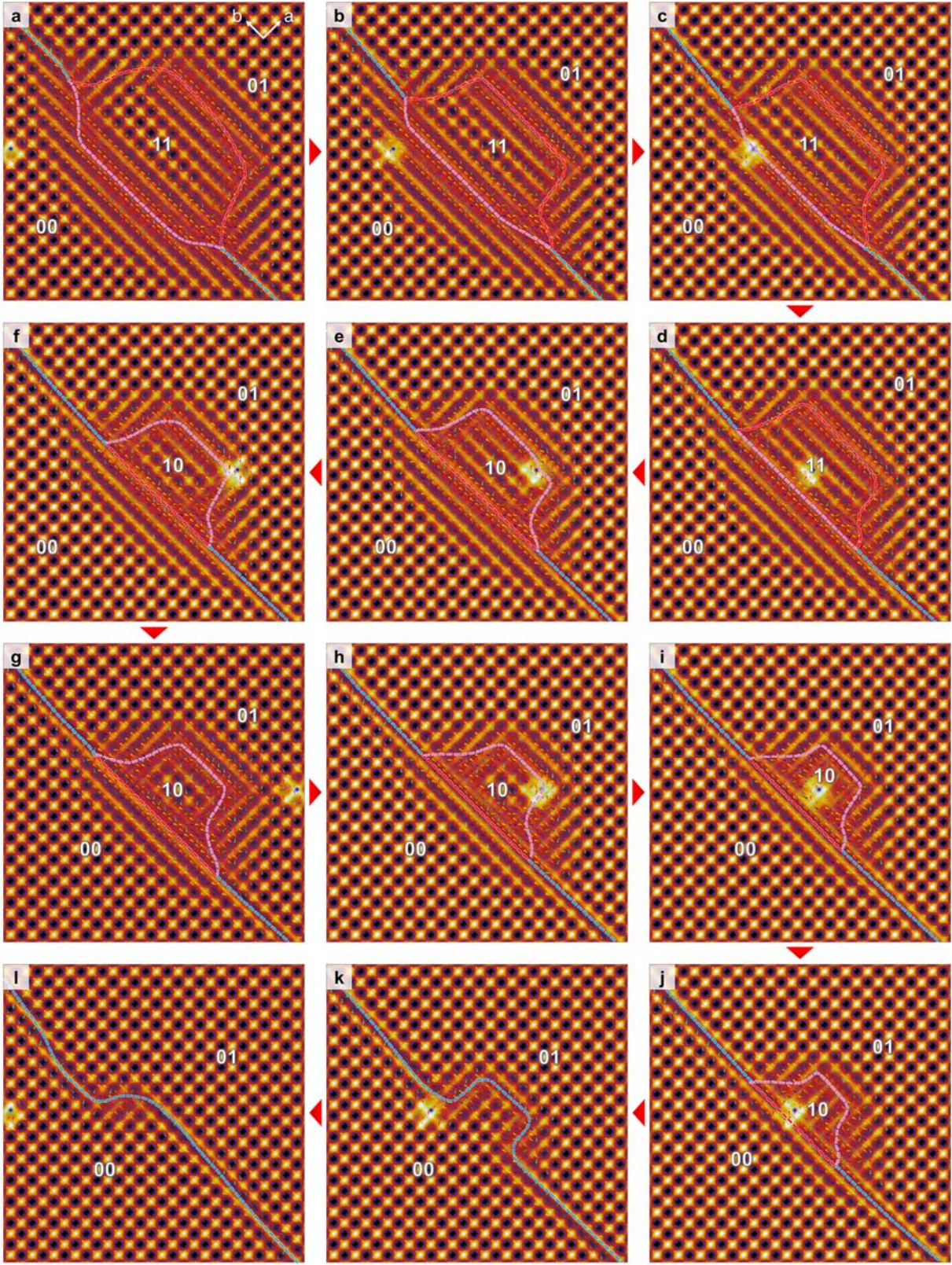

**Figure S15.** Higher spatial- and time-resolution images for simulation in Fig. 5 i-m, showing the changes in spin configurations near the complex pDWs in detail. Sudden phase changes of localized domain island is visible in **e** and **k** where the latter case eliminates a pDW and frees the localized domain island.

**SI References**

36. Caron, J. M., Neilson, J. R., Miller, D. C., Llobet, A. and McQueen, T. M. Iron displacements and magnetoelastic coupling in the antiferromagnetic spin-ladder compound $BaFe_2Se_3$. *Phys. Rev. B* **84**, 180409(R) (2011).

37. Bode, M., Vedmedenko, E. Y., von Bergmann, K., Kubetzka, A., Ferriani, P., Heinze, S., and Wiesendanger, R. Atomic spin structure of antiferromagnetic domain walls. *Nat. Mater.* **5**, 477 - 481 (2006).

38. Massee, F. *et al.* Cleavage surfaces of the $BaFe_{2-x}Co_xAs_2$ and $Fe_ySe_{1-x}Te_x$ superconductors: A combined STM plus LEED study. *Phys. Rev. B* **80**, 140507 (2009).